\RequirePackage{etex}
\documentclass[12pt]{article}
\usepackage{amsmath}
\usepackage{graphicx,psfrag,epsf}
\usepackage{enumerate}
\usepackage{lipsum} 
\usepackage[utf8]{inputenc}
\usepackage[style=ieee, citestyle=numeric-comp, backend=biber]{biblatex}

\usepackage{url} 
\usepackage[T1]{fontenc} 
\usepackage{mathtools} 
\usepackage[lofdepth,lotdepth]{subfig} 
\usepackage{mwe} 
\usepackage{amssymb}
\usepackage{hyperref}
\usepackage{xcolor}
\usepackage{adjustbox}
\usepackage{multirow}
\usepackage{pdflscape}
\usepackage[final]{pdfpages}
\usepackage[mathscr]{euscript}
\usepackage{etex}
\usepackage{amsthm}
\usepackage{tabulary}
\usepackage{varwidth}
\usepackage{caption} 
\usepackage{tikz}
\usepackage{pgfplots}  
\usepackage{array} 
\usepackage{geometry}
\usepackage{enumitem}
\pgfplotsset{compat=1.16} 
\makeatletter
\renewcommand{\eqref}[1]{Eq.~(\ref{#1})}
\newcommand*{\rom}[1]{\expandafter\@slowromancap\romannumeral #1@}
\makeatother

\newcommand{\Rv}{\boldsymbol{R}}

\newcommand{\zv}{\boldsymbol{z}}
\newcommand{\Zv}{\boldsymbol{Z}}

\newcommand{\dv}{\boldsymbol{d}}
\newcommand{\muv}{\boldsymbol{\mu}}
\newcommand{\nullv}{\boldsymbol{0}}
\newcommand{\Sigmav}{\boldsymbol{\Sigma}}

\definecolor{ao(english)}{rgb}{0.0, 0.5, 0.0}

\begin{document}



\title{Efficiency of Multivariate Tests in Trials in Progressive Supranuclear Palsy}

\author{Elham Yousefi\textsuperscript{1} \and Mohamed Gewily\textsuperscript{2} \and Franz K\"onig\textsuperscript{1}   \and G\"unter H\"oglinger\textsuperscript{3,4,5,6} \and Franziska Hopfner\textsuperscript{3} \and Mats O. Karlsson\textsuperscript{2} \and Robin Ristl\textsuperscript{1} \and Sonja Zehetmayer\textsuperscript{1} \and Martin Posch\textsuperscript{1,}\footnote{Correspondence to: martin.posch@meduniwien.ac.at. \textsuperscript{1}Center for Medical Data Science, Medical University of Vienna, Vienna, Austria. \textsuperscript{2}Department of Pharmacy, Uppsala University, Uppsala, Sweden. \textsuperscript{3}Department of Neurology, LMU University Hospital, Ludwig-Maximilians-Universität (LMU) München, Munich, Germany. \textsuperscript{4}German Center for Neurodegenerative Diseases (DZNE), Munich, Germany. \textsuperscript{5}Munich Cluster for Systems Neurology (SyNergy), Munich, Germany. \textsuperscript{6}Department of Neurology, Hannover Medical School, Hanover, Germany.}}
\date{\today}
\maketitle

\begin{abstract}

Objective: Measuring disease progression in clinical trials for testing novel treatments for multifaceted diseases as Progressive Supranuclear Palsy (PSP), remains challenging. In this study we assess a range of statistical approaches to compare outcomes measured by the items of the Progressive Supranuclear Palsy Rating Scale (PSPRS).

Methods: We consider several statistical approaches, including sum scores, as an FDA-recommended version of the Progressive Supranuclear Palsy Rating Scale (PSPRS), multivariate tests, and analysis approaches based on multiple comparisons of the individual items. In addition, we propose two novel approaches which measure disease status based on Item Response Theory models. We assess the performance of these tests under various scenarios in an extensive simulation study and illustrate their use with a re-analysis of the ABBV-8E12 clinical trial. Furthermore, we discuss the impact of the FDA-recommended scoring of item scores on the power of the statistical tests.

Findings: We find that classical approaches as the PSPRS sum score demonstrate moderate to high power when treatment effects are consistent across the individual items. The tests based on Item Response Theory (IRT) models yield the highest power when the simulated data are generated from an IRT model. The multiple testing based approaches have a higher power in settings where the treatment effect is limited to certain domains or items. The FDA-recommended item rescoring tends to decrease the simulated power in most test settings.

Conclusions: The study demonstrates that there is no one-size-fits-all testing procedure for evaluating treatment effects using PSPRS items; the optimal method varies based on the specific effect size patterns. The efficiency of the PSPRS sum score, while generally robust and straightforward to apply, varies depending on the specific patterns of effect sizes encountered and more powerful alternatives are available in specific settings. These findings can have important implications for the design of future clinical trials in PSP and similar multifaceted diseases.
\end{abstract}

\linespread{1.6}
\newpage
\section{Introduction} \label{sec-Introduction}
In diseases that exhibit multifaceted manifestations, disease progression cannot be characterised with a single measurement. Instead, multiple characteristics have to be assessed to describe the disease status \cite{FDAReport2017}. This poses a challenge to define appropriate endpoints in clinical trials to assess the effect of investigational treatments on the disease progression. 

In this paper, we assess a wide range of statistical approaches that have been proposed to define and analyse endpoints for clinical trials in indications where disease progression is measured by several variables. We focus on the setting of clinical trials for
Progressive Supranuclear Palsy (PSP), a rare neurodegenerative disorder with complex symptoms that affect balance, vision, body movements, and speech, ultimately leading to death. The most commonly used endpoint is the Progressive Supranuclear Palsy Rating Scale (PSPRS), a 28-item method to measure progression of the disease \cite{golbe2007clinical}. The item scores are categorical values which mostly vary from zero to 4. Recently, the FDA suggested to use only a subset of 10 item and to rescore some of the items into coarser categories (see Section \ref{subsec-ProgressiveSupranuclearPalsyRatingScale} below). The PSPRS is then defined as a sum score, summing the item scores either of all 28-items (for the original score) or the subset of the 10 items suggested by FDA.

The use of a sum score, such as the PSPRS or its modifications, is a standard approach to combining information from multiple endpoints. By using such an aggregated score, standard statistical tests like an analysis of covariance can be applied to compare the outcome variable, or its change from baseline, between groups.

Another approach to demonstrate efficacy with an overall test involves multivariate testing procedures\cite{o1984procedures,reitmeir1996one}. 
These procedures aggregate test statistics of the comparisons of the individual characteristics into a single univariate test statistic and test the global null hypothesis that the treatments do not differ in any of the individual endpoints. \citeauthor{o1984procedures} proposed two global directional tests known as Ordinary Least Squares (OLS) and Generalised Least Squares (GLS) tests to demonstrate an overall treatment effect \cite{o1984procedures}. Global directional tests are proposed in settings, where treatment benefit corresponds to a change into the same direction in all item scores. In such settings, two-sided tests as the Hotelling’s $T^2$ test, which does not account for the direction of treatment differences, is not of interest and will therefore not be considered further \cite{ristl2019methods}.

An alternative approach is to consider the individual components of the multivariate endpoint separately. In this case, inference is based on the individual test statistics or the individual p-values for the multiple endpoints and a multiplicity correction is applied to control the familywise error rate in the strong sense. This can be achieved by the Bonferroni procedure or more powerful multiple tests that take the dependence between the endpoints into account.

A further univariate test we consider is based on an Item Response Theory (IRT) model \cite{rasch1960studies}. As endpoint we consider for each patient the value of the latent variable of the IRT model, which describes the disease status and can be predicted based on estimating the IRT model parameters given the data (or the scores). As the estimation of the latent variable from the item scores is complex we propose also to approximate the latent variable estimate using  linear models. Specifically, we approximate the latent variable by a weighted sum of the item scores. This linear model-based estimation makes the endpoint better interpretable and yields a similar endpoint as the PSPRS sum score, however, using a weighted sum with weights derived from the IRT model (see Section \ref{sec-Analysis.methods}).

We aim to identify the most powerful statistical tests to assess treatment effects under various scenarios. To compare the power of the different procedures, we perform a comprehensive simulation study. Especially, we consider different strategies to simulate data of the PSPRS item scores, including the discretisation of multivariate normal outcomes as well as resampling from actual clinical trial data from the ABBV-8E12 trial \cite{hoglinger2021AbbVie}. In addition, we consider a range of alternative hypotheses to cover settings where there is a treatment effect in all item scores or only in a subdomain or even in a single item score only. Additionally, we investigate the impact of the scoring of the items, suggested by the FDA, on the power of the considered statistical methods, as well as the case where the original scores are considered (See Section \ref{subsec-ProgressiveSupranuclearPalsyRatingScale}).

This paper is structured as follows: Section \ref{subsec-ProgressiveSupranuclearPalsyRatingScale} describes the Progressive Supranuclear Palsy Rating Scale and its FDA modification. Section \ref{Section:Case_Study} discusses the ABBV-8E12 trial, forming the basis of our analysis. The considered analysis methods are detailed in Section \ref{sec-Analysis.methods}, while Section \ref{sec-Simulations} describes the different simulation approaches for the item level data as well as the simulation results. A re-analysis of the ABBV-8E12 trial is provided in Section \ref{sec-Illustration}. The paper concludes with a discussion in Section \ref{sec-Discussion}. Technical details on multivariate tests are provided in Section~1 of Supplementary Material. 
\section{Case Study\label{Section:Case_Study}}

As case study to inform the trial designs and simulation of trial data in the simulation study, we considered  the ABBV-8E12 trial \cite{hoglinger2021AbbVie}. The ABBV-8E12 trial was a randomised three-armed parallel group trial comparing a 2000mg and 4000mg dose of the investigational compound Tilavonemab to placebo.

In the ABBV-8E12 trial a total of 377 patients received at least one dose of the investigational treatment or placebo (Tilavonemab 2000 mg $n=126$, Tilavonemab 4000 mg $n=125$, placebo $n=126$). The original PSPRS items were assessed at baseline and weeks 12, 24, 36 and 52 where the change from baseline to week 52 in the PSPRS total score was the main outcome variable. In order to assess efficacy of treatments, the change from baseline to week 52 was analysed using a mixed-effect repeated measure model. The Bonferroni approach was used to account for multiplicity due to multiple comparisons between the two doses and placebo groups. The change from baseline to week 52 in PSPRS total score was similar between the three treatment groups at all considered visits and no significance difference between treatment groups, in favor of Tilavonemab 2000 mg or Tilavonemab 4000 mg versus placebo, at the two-sided significance level of 5\% was observed.

To investigate the different testing approaches, below we considered a simplified study design with only one treatment-control comparison. 
\section{Methods}\label{sec-Methods}
\subsection{The Progressive Supranuclear Palsy Rating Scale (PSPRS)}\label{subsec-ProgressiveSupranuclearPalsyRatingScale}
Multiple clinical trials have been conducted to investigate the efficacy of therapeutic interventions in Progressive Supranuclear Palsy \cite{hoglinger2021AbbVie,dam2021Biogen,boxer2014davunetide,nuebling2016prospera,tolosa2014Tideglusib}. However, so far no symptomatic treatments or disease-modifying therapies are available \cite{roesler2019four,levin2016differential}.

In randomised, phase 2, placebo-controlled trials of experimental treatments of PSP, the standard method to test for a treatment effect is to use the sum of the 28 item scores (the so called total score) of the Progressive Supranuclear Palsy Rating Scale (PSPRS) as the primary endpoint variable. Changes in this endpoint are compared between treatment groups, with higher scores indicating more severe disease conditions. Recently, the FDA has recommended to use a modified version of the PSPRS, which includes a subset of 10 items (Table \ref{table.FDAsubsetofPSPRS.fullname,shortName}). Each item in this subset is assigned a score ranging from zero to 4. 
Note that the score is based on the physicians' assessment  and is measured repeatedly to assess disease progression across different visits in the trials.

In addition, FDA recommended to collapse some of the item levels of the 10-item version of the PSPRS. A re-scoring has been recommended by FDA for all but two items, since it was argued that the original response levels for some items do not reflect clear and clinically meaningful differences of the patients' conditions and are therefore difficult to interpret. Throughout the remainder of the text, we refer to the original and rescored item levels as the "original scores" and the "FDA scores". {The full version of the PSPRS (with 28 items) which includes the subset of 10 items suggested by the FDA is provided as a table in the online Supplement B. The table contains the original scores and the FDA scores, including a description on how to collapse the item levels.}

\begin{table}[htp]
    \caption{FDA-recommended 10-item version of the PSPRS. The 10 items are divided into 3 domains (History, Bulbar exam, Gait/Midline Exam). The numbering of the items corresponds to the original PSPRS score \cite{golbe2007clinical}. The column "Abbreviation" gives the short name used to label the items in the subsequent graphs.}
    \label{table.FDAsubsetofPSPRS.fullname,shortName}
    \centering
    \resizebox{\textwidth}{!}{
        \begin{small}
            \begin{tabular}{p{0.8\textwidth}p{0.2\textwidth}}
                \hline \hline
                \textbf{Full name} & \textbf{Abbreviation} \\
                \hline \hline
                \textbf{\rom{1}. HISTORY (from patient or other informant)} &\\
                3. Dysphagia for solids (from patient or other informant) & Dysp.FS \\
                4. Using knife and fork, buttoning clothes, washing hands and face (rate the worst) & Use.KF \\
                5. Falls (average frequency if patient attempted to walk unaided) & Fall \\
                \hline
                \textbf{\rom{3}. BULBAR EXAM} &\\
                12. Dysarthria (ignoring palilalia) & Dysa. \\
                13.  Dysphagia (for 30-50 cc of water from a cup, if safe) & Dysp. \\
                \hline
                \textbf{\rom{6}. GAIT/MIDLINE EXAM} &\\
                24. Neck rigidity or dystonia & Neck.Ri \\
                25. Arising from chair & Ari.FC \\
                26. Gait & Gait \\
                27. Postural stability (on backward pull) & Pos.St \\
                28. Sitting down (may touch seat or back but not arms of chair) & Sit \\
                \hline \hline
            \end{tabular}
        \end{small}
    }
\end{table}

In this work we focus on the item scores of the 10 items proposed by FDA. This is also supported by a statistical assessment of the 10 items based on IRT analysis, which  also suggests that these items are the most informative to describe the course of the disease \cite{gewily2023irt}.

Additionally, as in the ABBV-8E12 trial \cite{hoglinger2021AbbVie}, we assume that the primary endpoint is based on the item scores measured at week 52, accounting for the baseline, corresponding to a one year observation period.

\subsection{Analysis methods} \label{sec-Analysis.methods}

All the analyses below were performed based on the subset of 10 FDA recommended items of the PSPRS with the original scoring as well as the FDA re-scoring.

With the exception of the IRT based test (but not the approximate IRT based test), all tests below test the null hypothesis that the expected values of all of the FDA recommended item scores are larger or equal in the treatment group than in the control group. We term this null hypothesis as the strong null hypothesis. The corresponding alternative hypothesis is that the expected values of at least one item scores is strictly smaller in the treatment group compared to the control group. Some of the tests in addition are valid tests for broader null hypotheses as indicated below.

\paragraph{PSPRS Scores}As benchmark, we used an analysis of covariance (ANCOVA) for the PSPRS sum score at week 52 as dependent variable and the PSPRS sum score at baseline as well as the treatment indicator as independent variables. The (t-)test for the coefficient for the treatment variable then tests the null hypothesis of no treatment effect. Note that this adjusted analysis is equivalent to a between group comparison of the change from baseline between the two groups.
This test not only tests the strong null hypothesis, but also the null hypothesis that the mean of the item scores (across items and the patient population) is larger or equal in the treatment compared to the control group. 

\paragraph{IRT-based test} The second considered, univariate test is based on estimates of the latent variable based on an IRT model \cite{siemons2014short}. IRT models represent the state of disease as a latent variable which is defined based on multiple reported measures. In PSP, the latent variable corresponding to the disease severity is measured through the PSPRS item questionnaire as reported by the physician. To perform the IRT-based test, we require an IRT model fitted on an external data set. We used a graded response (GR) model,  fitted to the ordered polytomous data \cite{ueckert2018modeling} scores of the FDA-recommended 10-item version of the PSPRS from the ABBV-8E12 trial. To this end, we first pooled the data across treatment groups and visits (including the baseline and follow-up visits at weeks  12, 24, 36
and 52) such that the data from each visit became an independent individual (row) in the data set. Based on this data set, we fitted an IRT model. The parameter estimates of the GR model are presented in Table~1 of Supplementary Material A. This simplified analysis does not account for the dependence of the measurements from a patient at different visits, but accounts for the dependence of a patient's measurements of the different items within a single visit. This allowed to estimate the IRT model based on a larger data set with the aim to enhance the precision of the model estimates.

Based on this a-priori estimated model, we then estimate the latent trait variable for the patients in the actual clinical trial based on the observed item scores, separately for the baseline and the 52-weeks data. Especially, we estimate the predicted latent trait variable with the expected a-posteriori (EAP) method \cite{thissen1995item}, which is implemented, for example, in the \texttt{R} package \texttt{mirt} for the analysis of dichotomous/polytomous response data using latent trait models under the IRT paradigm \cite{chalmers2012mirt}. 

Finally, an ANCOVA model is fitted with the predicted latent trait variable at week 52 as dependent and the latent trait variable at baseline and treatment as independent variables. This procedure tests the null hypothesis that the expected value of the latent variables is larger in the treatment group than the control group.

This procedure tests the null hypothesis that the expected values of the latent variable estimate in the treatment group is larger or equal than in the control group. Especially, this null hypothesis holds, if the joint distribution of item scores in the two treatment groups is equal. However, given the complex dependence of the latent variable estimates on the item scores, it is not obvious if this null hypothesis is included in the strong null hypothesis defined above.

\paragraph{Approximate IRT-based test }
As the estimation of the latent variable with IRT model is complex and the impact of the individual item scores
is not immediately understandable without going through complex derivations, we aim to approximate the estimate by a weighted sum of the item scores. This aims to make the endpoint better interpretable. Therefore, we consider as endpoint a weighted sum of the item scores, where the weights are derived from a linear model fitted on the data of an external trial. We used again data of the ABBV-8E12  trial, where, as above, we aggregated the data from all treatment groups and visits, fitted and IRT model and estimated the latent variable for each patient. Then, using the same data set, we fitted a linear model with the latent variable as dependent variable and the 10 FDA recommended items as independent variable. Note that the latent variable was transformed with the standard logistic function to improve model fit.
Based on this a-priori estimated model, we then estimate the (approximate) latent trait variable for the patients in the actual clinical trial by computing the weighted sum of the observed item scores and back transforming it (using the standard logit function) to match the original scale. Note that in principle the linear model could give values outside the unit interval such that the standard logit function cannot be applied. If this occurs one can truncate the values accordingly. For our data set this was not an issue as the corresponding fitted linear model only gives values in the unit interval. In addition to the strong null hypothesis defined above, this procedure tests the null hypothesis that the expected value of the resulting weighted sum score is larger in the treatment compared to the control group.

\paragraph{OLS and GLS tests}
The \citeauthor{o1984procedures}’s OLS and GLS tests \cite{o1984procedures} are multivariate tests combining test statistics of separate tests comparing the responses for each of the items between groups to a common test statistics. 
The OLS test is based on the standardised,  unweighted sum of t-statistics of the individual items,  see Eq.~(1) in Supplementary Material A. The GLS test, in contrast, is based on a weighted sum, determined by the row sums of the inverse of the correlation matrix (see  Eq.~(2) in Supplementary Material A). The t-statistics for the comparison of the individual item scores are obtained from an ANCOVA model, with the item score at week 52 as the dependent variable and the baseline item score and treatment indicator as independent variables. The correlation of the test statistics, required to standardize the test statistics and, for the GLS test, to compute the weights, is estimated using multiple marginal modeling (using the mmm function \cite{pipper2012versatile} in the \texttt{multcomp} package in \texttt{R}). Endpoints that are highly correlated receive lower weights in the GLS test statistic. For very high correlations, the weights may even become negative. This implies that the corresponding test may no longer be a directional test but can inflate the type I error rate if there is a negative effect in some of the items and no effect in the others. Thus, the GLS test controls the type 1 error rate under the strong null hypothesis only if all weights are non-negative. When re-analysing the  ABBV-8E12 trial, we observed a negative weight for one item and therefore considered a modified GLS test, where we dropped this item (see Section \ref{sec-Illustration}). Following the recommendation of \cite{logan2004brien}, for the OLS and GLs tests we used a modified approximation of the degrees of freedom (df) setting  $df=0.5(2 n-3)(1+1/m^2)$, where $n$ is the per-group sample size and $m$ the number of items. This has been shown to provide better control of the type I error rate for smaller sample sizes (see Supplementary Material A, Subsection~1.1) compared to the original proposal by O'Brien.

\paragraph{Bonferroni test} An alternative approach to computing cumulative test statistics, based on individual tests for each item (as the OLS or GLS test), is to consider the results of each of the individual tests separately and to apply a multiplicity adjustment to control the FWER. Individual t-statistics (and the associated un-adjusted p-values) are obtained from multiple marginal models as described above. Then an adjusted significance level is applied. The overall null hypothesis of no treatment effect in any item, is rejected if the smallest p-value across all items falls below the Bonferroni adjusted significant level (which is $\alpha/10$, as 10 items are considered). An advantage of this approach is that it also provides a test to compare the item scores for each item: One can reject the null hypothesis of no treatment effect for an item, if the corresponding p-value falls below the Bonferroni adjusted significance level. A further improvement for the individual tests can be obtained by applying the Bonferroni-Holm test. 

\paragraph{Hommel/Simes test} Similar to the Bonferroni test, also other multiple testing procedures can be applied to test the overall null hypothesis. An example is the  Simes \cite{simes1986improved} test for the global null hypothesis of no treatment effect in any item (and its closure the Hommel test \cite{hommel1988stagewise}, to obtain tests for the individual items). This test is, for example in the \texttt{R} package \texttt{hommel}. For these tests control of the type I error rate has been shown for independent test statistics as well as test statistics with certain positive correlation structures.

\paragraph{Omnibus test} The Omnibus test \cite{futschik2019omnibus} is an alternative test for the global null hypothesis of no treatment effect in any item  which is based on individual test statistics. This test has shown to have robust power both in settings where for a few or for many null hypotheses the alternative hypothesis holds false. It is based on cumulative sums of (reciprocal) transformed sorted p-values. Note that theoretical type I error rate control for these tests has been shown for independent test statistics only.

\paragraph{Omnibus test for domain scores} 
The 10 items of the PSPRS scale are divided into three domains (see Table \ref{table.FDAsubsetofPSPRS.fullname,shortName}). 
For this test we first compute the sum-score  across items for each domain. We then perform the ANCOVA analysis for each of the resulting three domain scores and then adjust for multiplicity using the Omnibus test as above.

\paragraph{MaxT test}
The Bonferroni approach is strictly conservative  as it does not take into account the correlation of test statistics. Based on the estimated correlation of t-statistics  one can use a test based on the individual t-values which takes into account the correlations. 
To improve the accuracy of the subsequent normal approximation, we first replace the t-values by transformed z-values to take into account the degrees of freedom, setting $z_i=\Phi^{-1}(F_{t,df}(-t_i)), i=1,\dots,10$, where  $\Phi^{-1}$ and $F_{t,df}$ denote respectively the quantile function of the normal distribution and the CDF of the t-distribution.  

Since low values of the item scores correspond to a better outcome, small (negative) t-values or equivalently large (positive) z-values, as defined above, indicate a beneficial treatment effect. Therefore, the test statistics to test the overall null hypothesis of no (beneficial) treatment effect in any item, is defined as the  maximum of the z-statistics. 
Then, the multiplicity adjusted p-value is obtained, based on the distribution function of the multivariate normal distribution and is given by  
$P_{\Zv_{\max}}=1-P_{\Sigmav}(\Zv\leq \zv_{\max})$, $\Zv \sim \mathcal{N}(\nullv,\,\hat{\Sigmav})\,$
where $\zv_{\max}=\left(z_{\max},\dots,z_{\max}\right)$, $z_{\max}=\max(z_1,\dots,z_{10})$ and the correlation matrix $\hat{\Rv}$ is estimated using multiple marginal models as above (see the paragraph on OLS and GLS tests).

The considered analysis methods are summarized in Table \ref{table.test.description}. 


\begin{table}[htp]
\linespread{1.6}
    \caption{The considered analysis methods: category of the method (first column),  name of the analysis method (second column), short name used to label the method in the result section (third column), and description for each test (fourth column). 
    }
 \label{table.test.description}
    \begin{center}
    \resizebox{\textwidth}{!}{
    \begin{tabular}{p{0.3\textwidth}p{0.15\textwidth}p{0.15\textwidth}p{0.4\textwidth}}
        \hline \hline
        \parbox[t]{0.3\textwidth}{\linespread{1}\selectfont Category}&\parbox[t]{0.15\textwidth}{\linespread{1}\selectfont Name}&\parbox[t]{0.15\textwidth}{\linespread{1}\selectfont Abbreviation} & \text{Description} \\
        \hline \hline
        \multirow{3}{*}{\parbox[t]{0.3\textwidth}{\linespread{1}\selectfont Univariate test statistic (of aggregated scores )}} & \parbox[t]{0.15\textwidth}{\linespread{1}\selectfont Sum Score}& \parbox[t]{0.15\textwidth}{\linespread{1}\selectfont SumS} & \parbox[t]{0.4\textwidth}{\linespread{1}\selectfont Sum of the item scores is used as the dependent variable.} \\
        \cline{2-4}
        & \parbox[t]{0.15\textwidth}{\linespread{1}\selectfont IRT-based test}& \parbox[t]{0.15\textwidth}{\linespread{1}\selectfont IRT.PSIF} & \parbox[t]{0.4\textwidth}{\linespread{1}\selectfont IRT-based estimate of the latent variable is used as the endpoint.} \\
         \cline{2-4}
        & \parbox[t]{0.15\linewidth}{\linespread{1}\selectfont Linear model-based test\\}& \parbox[t]{0.15\textwidth}{\linespread{1}\selectfont LM.PSIBPF} & \parbox[t]{0.4\textwidth}{\linespread{1}\selectfont uses a linear model-based estimate of the latent variable, with a weighted sum of the item scores, as the endpoint.} \\
         \hline \hline
        \multirow{3}{*}{\parbox[t]{0.3\textwidth}{\linespread{1}\selectfont Weighted test statistic composed of that of the individual items.}}& \parbox[t]{0.15\textwidth}{\linespread{1}\selectfont O’Brien’s OLS test}  & \parbox[t]{0.15\textwidth}{\linespread{1}\selectfont OLS} & \parbox[t]{0.4\textwidth}{\linespread{1}\selectfont sum of t-statistics of the individual items with equal weights (see Supplementary Material A, Subsection~1.1).} \\
        \cline{2-4}
         & \parbox[t]{0.15\textwidth}{\linespread{1}\selectfont O’Brien’s GLS test}& \parbox[t]{0.15\textwidth}{\linespread{1}\selectfont GLS}  & \parbox[t]{0.4\textwidth}{\linespread{1}\selectfont sum of t-statistics of the individual items with unequal weights (see Supplementary Material A, Subsection~1.1).} \\
         \cline{2-4}
          & \parbox[t]{0.15\textwidth}{\linespread{1}\selectfont O’Brien’s GLS test}& \parbox[t]{0.15\textwidth}{\linespread{1}\selectfont GLS-26} & \parbox[t]{0.4\textwidth}{\linespread{1}\selectfont the GLS test based on 9 items (after eliminating item 26).} \\
        \hline \hline
        \multirow{5}{*}{\parbox[t]{0.3\textwidth}{\linespread{1}\selectfont Individual (item) test statistic}} & \parbox[t]{0.15\textwidth}{\linespread{1}\selectfont Bonferroni correction\\} & \parbox[t]{0.15\textwidth}{\linespread{1}\selectfont Bonf} & \parbox[t]{0.4\textwidth}{\linespread{1}\selectfont based on the smallest p-value across the individual test statistics of the 10 items.} \\
        \cline{2-4}
         & \parbox[t]{0.15\textwidth}{\linespread{1}\selectfont maximum T-value} & \parbox[t]{0.15\textwidth}{\linespread{1}\selectfont MaxT} & \parbox[t]{0.4\textwidth}{\linespread{1}\selectfont based on the largest (transformed) t-value among the 10 items.} \\
         \cline{2-4}
         & \parbox[t]{0.15\textwidth}{\linespread{1}\selectfont Simes} & \parbox[t]{0.15\textwidth}{\linespread{1}\selectfont Simes}  & \parbox[t]{0.4\textwidth}{\linespread{1}\selectfont based on individual adjusted p-values from the items.} \\
         \cline{2-4}
        & \parbox[t]{0.15\textwidth}{\linespread{1}\selectfont Omnibus}& \parbox[t]{0.15\textwidth}{\linespread{1}\selectfont Omnibus} & \parbox[t]{0.4\textwidth}{\linespread{1}\selectfont based on
cumulative sums of the transformed sorted p-values when there is no a priori knowledge on the number of false individual null hypotheses.} \\
         \cline{2-4}
         & \parbox[t]{0.16\textwidth}{\linespread{1}\selectfont Combination test: Omnibus and Sum Score\\} & \parbox[t]{0.14\textwidth}{Omnibus-dom} & \parbox[t]{0.4\textwidth}{\linespread{1}\selectfont a combined test applying the sum score test within each of the domains and the Omnibus test between them.} \\
        \hline \hline
    \end{tabular}
   }
  \end{center}
   \linespread{1.6}
\end{table}

\section{Simulation Study}\label{sec-Simulations}
To evaluate the operating characteristics of the considered analysis methods,
we conducted a large scale simulation study using three approaches to simulate individual item scores of the FDA-recommended subset of items of the PSPRS. This allows us to assess the robustness of findings with respect to specific assumptions of the data generating process. Especially, we generated data (1) from a discretised multivariate normal distribution, (2) with a Bootstrap approach based on  the ABBV-8E12 study, and (3) based on a longitudinal IRT model. As no differences between groups were observed,  we used pooled (across  treatment groups)  estimates of the  outcomes distribution and also pooled the groups to generate the Bootstrap samples (see the description below).

 \subsection{Simulation of the item level data\label{subsection:simulation_item_level_data}}
 
Below, we describe the three data simulation approaches in detail.  Table \ref{table.pars_simulation.description} summarizes the key aspects of the different approaches. 

All the simulation methods yielded scores using the original scoring of the items. To generate items according to the FDA re-scoring, we collapsed in a further step the  corresponding item levels.
In the  simulations we considered a trial with an experimental treatment and a control arm, assuming a per-group sample size of $n=70$. For each scenario $10.000$ trials were simulated.

\begin{table}[htp]
	\caption{Parameter specifications for simulation studies.}
	\label{table.pars_simulation.description}
    \begin{center}
    \resizebox{\textwidth}{!}{
			
				\begin{tabular}{p{0.2\textwidth}p{0.12\textwidth}p{0.3\textwidth}p{0.4\textwidth}}
 \hline \hline
\text{Name} &\text{Type}&\parbox[t]{0.2\textwidth}{\linespread{1}\selectfont Values} & \text{Description} \\
\hline \hline

\text{Analysis type}&\parbox[t]{0.1\textwidth}{\linespread{1}\selectfont Design choice}&\text{two covariates}&\parbox[t]{0.4\textwidth}{\linespread{1}\selectfont \linespread{1}\selectfont Analysis is based on ANCOVA models including the treatment effect and the relevant baseline value, to the analysis methods summarized in Table \ref{table.test.description}, as covariates.\\}\\
\hline

\parbox[t]{0.2\textwidth}{\linespread{1}\selectfont Total sample size per treatment}&\parbox[t]{0.1\textwidth}{\linespread{1}\selectfont Design choice}&\parbox[t]{0.3\textwidth}{\linespread{1}\selectfont 70 per treatment group}&\parbox[t]{0.4\textwidth}{\linespread{1}\selectfont The sample size is fixed across simulation approaches.}\\
\hline

\parbox[t]{0.2\textwidth}{\linespread{1}\selectfont Number of simulations runs}&\parbox[t]{0.12\textwidth}{Assumption}&\parbox[t]{0.3\textwidth}{\linespread{1}\selectfont 10000 per simulation approach}&\parbox[t]{0.4\textwidth}{\linespread{1}\selectfont The number of simulations runs is fixed across simulation approaches.}\\
\hline

\parbox[t]{0.2\textwidth}{\linespread{1}\selectfont Nominal significance level}&\parbox[t]{0.12\textwidth}{Assumption}&\parbox[t]{0.3\textwidth}{\linespread{1}\selectfont $\alpha=0.025$ }&\parbox[t]{0.4\textwidth}{\linespread{1}\selectfont The one-sided nominal significance level is fixed in all analysis methods and across simulation approaches.}\\
\hline

\parbox[t]{0.2\textwidth}{\linespread{1}\selectfont Parametric simulation method: discretised item scores}&\parbox[t]{0.12\textwidth}{Assumption}&\parbox[t]{0.3\textwidth}{\linespread{1}\selectfont effect sizes are either fixed/varied across all items or fixed across domains (details in Table \ref{table.delta.power.d2.5,baselin})}&\parbox[t]{0.4\textwidth}{\linespread{1}\selectfont First, data is generated from the multivariate normal distribution (see details in Subsection \ref{subsection:simulation_item_level_data}). It is then discretised to the nearest integer (and bounded from 0-4) to perform the analysis tests.}\\
\hline

\parbox[t]{0.2\textwidth}{\linespread{1}\selectfont Non-parametric simulation approach: Bootstrap (without replacement)}&\parbox[t]{0.12\textwidth}{Assumption}&\parbox[t]{0.3\textwidth}{\linespread{1}\selectfont effect sizes are either fixed/varied across all items or fixed across domains (details in Table \ref{table.delta.power.d2.5,baselin})}&\parbox[t]{0.4\textwidth}{\linespread{1}\selectfont Data sets are resampled based  from the ABBV-8E12 study at baseline and Week 52 (Section \ref{sec-Methods})}\\
\hline

\parbox[t]{0.2\textwidth}{\linespread{1}\selectfont IRT-based simulation method}&\parbox[t]{0.12\textwidth}{Assumption}&\parbox[t]{0.3\textwidth}{\linespread{1}\selectfont effect sizes are fixed in all items. Multiple ratios of the progression equation, $\rho$ values (details in Subsection \ref{subsection:simulation_IRT based simulation}), are investigated as $0.45$, $0.50$, $0.55$, $0.60$, $0.65$, $0.70$, $0.75$.}&\parbox[t]{0.4\textwidth}{\linespread{1}\selectfont \linespread{1}\selectfont Subjects are sampled from the ABBV-8E12 data. Subsequently, corresponding relevant covariates are used to construct the progression equation and generate the IRT-based item level data (details in Subsection (Section \ref{sec-Methods}).}\\

					\hline
				\end{tabular}
			
   }
  \end{center}
\end{table}

\paragraph{Discretised multivariate normal distribution}
To parameterize the simulation distribution, we estimated the  $20\times 1$ mean vector  and $20\times 20$ covariance matrix of the item scores at baseline and Week 52 from the ABBV-8E12 trial data. Especially, 
the control mean was estimated based on the data of the placebo group and the co-variances were estimated in each of the three treatment groups and then a pooled estimate, $\hat{\Sigmav}_p$, was computed, weighting according to the sample sizes. 

Then, to generate the control group data, we drew vectors from a multivariate normal distribution with mean $\hat{\muv}_3$ and covariance $\hat{\Sigmav}_p)$. Similarly, the  experimental treatment groups data was drawn from a multivariate normal distribution with the same covariance matrix but mean $\hat{\muv}_3-\dv$, where $\dv$ denotes the vector of effect sizes across items. 

The vectors sampled from the normal distributions are then discretised by rounding, and trimming the scores to the range 0 to 4. 

\paragraph{Bootstrap}  
We resampled subjects from the ABBV-8E12 study and considered both, resampling with and without replacement, but report here only the results of the latter as both provided similar results. Resampling was performed from the complete cases (where all items at the baseline and Week 52 visit were available) pooled across all three treatment groups.
Then subjects of the control group and the experimental treatment group were sampled from this data set. For simulations under the alternative hypothesis, for the subjects sampled in the experimental treatment arms the item scores at Week 52 were adjusted by the assumed treatment effect. Especially, for an effect size $d_k$ and item scores $y_{k,i}$ for item $ k=1,...,10$ and subject $i=1,\ldots, n$ in the experimental treatment group, where $n$ is the per group sample size of the simulated sample, we adjust the scores $y_{k,i}$ such that the mean effect in the population (up the floor effects due to the bounded scale) is approximately $d_k$. This is achieved setting 
           \begin{enumerate}
           \item For all $k=1,\ldots,10$ and $i=1,\ldots,n$, set $y_{k,i}\to y_{k,i}-\lfloor d_k \rfloor,$ where $\lfloor d_k \rfloor$ denotes the floor value of the individual item effect size $d_k$.
		   \item Randomly select a subset of $np_k$ (rounded to the nearest integer) subject indices, denoted by $I_k$, where $p_k=d_k-\lfloor d_k \rfloor$ and set 
			 $y_{k,j}\to y_{k,j}-1$ for all $ j\in I$.
			\item Finally, trim the scores to the range of the scale, setting
    $y_{k,i}\to max(0,y_{k,i})$ for all $k=1,\ldots,10$ and $i=1,\ldots,n$.
		\end{enumerate}

\paragraph{IRT model based data generation}\label{subsection:simulation_IRT based 
simulation}
We simulated item level data for the baseline and Week 52 item scores based on the longitudinal IRT model of \cite{gewily2023irt} fitted on multiple data sets from a set of interventional trials in PSP.
They used a 2-parameter graded response (GR) model (see, e.g., Equation 4 in \cite{ueckert2018modeling}) based on discrimination and difficulty as item characteristic parameters. The GR model was fitted to the ordered polytomous data, scores of the FDA-selected subset of items (but ignoring the rescoring recommendation) from the PSPRS in all interventional studies.  The underlying latent variable at time $t$, in years, is represented through a disease progression model for individual $i$, given by 

$$\psi_{i}(t)=\psi_{i}(0)+ s_i t$$

where $\psi_{i}(0)$ and $s_i$ respectively denote the intercept and slope.  For a subject $i$, intercept and slope are given as a function of relevant baseline and slope covariates of the patient, as age, sex, and PSP diagnostic phenotype (see \cite{gewily2023irt} for a detailed description of the derivation of the corresponding prediction model). 
To model disease progression under the alternative, for  patients in the treatment group, we assumed a smaller slope in the longitudinal latent variable model. Especially, we assumed that the latent variable evolves as $\psi_{i}(0)+ \rho s_it$ , where $\rho<1$ indicates a beneficial treatment effect.

To simulate the data we sampled subjects from the  ABBV-8E12 study and based on the covariate information of the sampled subjects, we computed for each subject the predicted latent variable at baseline and Week 52. Based on these latent variables we then simulate the individual item scores using the items discrimination and difficulty parameter as estimated in \cite{gewily2023irt}.

\subsection{Considered treatment effect scenarios}
The performance of the different testing approaches depends on the pattern of treatment effects across the items. We therefore considered a range of treatment effect patterns. For the simulations based on discretised multivariate normal samples and the Bootstrap approach, the considered effect size scenarios are given in Table \ref{table.delta.power.d2.5,baselin} and include the case of small, intermediate, and 
strong equal absolute effects across items ($\dv_{1}^\prime,\dv_{2}^\prime,\dv_{3}^\prime$). Here, $\dv_{l}^\prime, l=1,\dots,12$ denote the transposed treatment effect vectors across items.
We further considered scenarios ($\dv_{4}^\prime -\dv_{9}^\prime)$ where there the treatment effect is limited to items in a single or two domains. Within each domain, the absolute treatment effects are assumed to be equal. For instance, in $\dv_{5}^\prime$ we assume an equal absolute treatment effect in each the FDA-recommended items in the History Domain (items 3, 4 and 5). As a reference, we have also included three scenarios in which the treatment effect is isolated to a single item ($\dv_{10}^\prime -\dv_{12}^\prime)$. 

For the data generated with the IRT model, we considered different effects on the slope of the latent variable, and specified $\rho$ as $\rho= \lbrace 0.45, 0.5,0.55,0.6,0.65,0.7,0.75 \rbrace$ which leads to power values that allow to discriminate results in the statistical tests with respect to power.

\begin{table}[htp]
		\caption{Effect size scenarios considered in the simulation study. The rows correspond to vectors $\dv$ of effect sizes for the 10 items in the FDA recommended modification of the PSPRS scale.}
		\label{table.delta.power.d2.5,baselin}
		\vspace{-0.5cm}
		\begin{center}
			\resizebox{\textwidth}{!}{
				\begin{scriptsize}
						\begin{tabular}{llcccccccccc}
							\hline
                            &&\text{Dysp.FS}& \text{Use.KF}& \text{Fall}& \text{Dysa}& \text{Dysp.}& \text{Neck.Ri}& \text{Ari.FC}& \text{Gait}& \text{Pos.St}& \text{Sit}\\
                            \hline
							\text{Equal effect size}&$\dv_{1}^\prime$&$0.20$&$ 0.20$&$ 0.20$&$ 0.20$&$ 0.20$&$ 0.20$&$ 0.20$&$ 0.20$&$ 0.20$&$  0.20$\\
                            \text{Equal effect size}&$\dv_{2}^\prime$&$0.25$&$ 0.25$&$ 0.25$&$ 0.25$&$ 0.25$&$ 0.25$&$ 0.25$&$ 0.25$&$ 0.25$&$  0.25$\\
                            \text{Equal effect size}&$\dv_{3}^\prime$&$0.30$&$ 0.30$&$ 0.30$&$ 0.30$&$ 0.30$&$ 0.30$&$ 0.30$&$ 0.30$&$ 0.30$&$  0.30$\\
							\text{History domain}&$\dv_{4}^\prime$&$0.85$&$ 0.85$&$ 0.85$&$ 0.00$&$ 0.00$&$ 0.00$&$ 0.00$&$ 0.00$&$ 0.00$&$  0.00$\\
							\text{Bulbar exam}&$\dv_{5}^\prime$&$0.00$&$ 0.00$&$ 0.00$&$ 1.25$&$ 1.25$&$ 0.00$&$ 0.00$&$ 0.00$&$ 0.00$&$  0.00$\\
							\text{Gait/midline exam}&$\dv_{6}^\prime$&$0.00$&$ 0.00$&$ 0.00$&$ 0.00$&$ 0.00$&$ 0.50$&$ 0.50$&$ 0.50$&$ 0.50$&$  0.50$\\
							\text{History domain \& Bulbar exam}&$\dv_{7}^\prime$&$0.50$&$ 0.50$&$ 0.50$&$ 0.50$&$ 0.50$&$ 0.00$&$ 0.00$&$ 0.00$&$ 0.00$&$  0.00$\\
							\text{History domain \& Gait/midline exam}&$\dv_{8}^\prime$&$0.30$&$ 0.30$&$ 0.30$&$ 0.00$&$ 0.00$&$ 0.30$&$ 0.30$&$ 0.30$&$ 0.30$&$  0.30$\\
							\text{Bulbar \& Gait/midline exam}&$\dv_{9}^\prime$&$0.00$&$ 0.00$&$ 0.00$&$ 0.35$&$ 0.35$&$ 0.35$&$ 0.35$&$ 0.35$&$ 0.35$&$  0.35$\\
							\text{Dysphagia for solids (in History domain)}&$\dv_{10}^\prime$&$2.50$&$ 0.00$&$ 0.00$&$ 0.00$&$ 0.00$&$ 0.00$&$ 0.00$&$ 0.00$&$ 0.00$&$  0.00$\\
							\text{Dysarthria (in Bulbar exam)}&$\dv_{11}^\prime$&$0.00$&$ 0.00$&$ 0.00$&$ 2.50$&$ 0.00$&$ 0.00$&$ 0.00$&$ 0.00$&$ 0.00$&$  0.00$\\
							\text{Neck rigidity (in Gait/midline exam)}&$\dv_{12}^\prime$&$0.00$&$ 0.00$&$ 0.00$&$ 0.00$&$ 0.00$&$ 2.50$&$ 0.00$&$ 0.00$&$ 0.00$&$  0.00$\\
							\hline
						\end{tabular}
				\end{scriptsize}
			}
		\end{center}
\end{table}

\subsection{Results of the simulation study}

In the simulation study, we assessed the type I error rate and power of each of the analysis methods for the different data generating procedures and treatment effect scenarios.

None of the considered methods showed an apparent inflation of the type I error rate in the considered scenarios (see Figure \ref{Figure: Alphasim,ALLSimulationApp,LMfit,5new-tests}). This holds for all three simulation approaches and the original item scoring as well as the tests based on the rescored items.

Figure \ref{Figure: Powersim,ALLSimulationApp,LMfit,5new-tests} summarizes the power of the different methods across all simulation approaches and effect size scenarios, both for the original scoring of the items as well as the FDA-rescoring. Based on 10000 simulations runs, the standard error of the power estimates is bounded from above by 0.005. To better distinguish the scenarios, separate plots for different treatment effect scenarios and data simulation approaches are given in the Figures 1 and 2 in the Supplementary Material A.

\paragraph{Comparison of testing procedures}
In treatment effect scenarios where there is a homogeneous absolute effect in all items (Scenarios $\dv_{1}^\prime$, $\dv_{2}^\prime$ and $\dv_{3}^\prime$) we observe the highest power for the test based on the classical PSPRS sum
score using the original scores. This holds for data simulated with Bootstrap as well as data simulated from the multivariate normal distribution. Similarly, the GLS test and its modification excluding item 26 (Gait) show high power values.  However, even after eliminating this item, we observed negative  weights for other items in  simulated data sets, which causes issues in interpretation.
In scenarios with homogeneous effects, also the test using the latent variable estimate from the IRT model as well as its approximation (with slightly lower power) and the Omnibus test based on the domain sum score show a good performance. In contrast, the other tests based on multiple testing procedures had substantially lower power.

On the other hand, when there is an effect only in certain domains or individual items, the power of tests based on multiplicity adjusted separate tests have the highest power. Here, the power of the test based on the PSPRS sum score is lower. The tests based on the IRT model have low power if there is an effect in a single domain or single item only. In considered scenarios where the effect is in two domains, the power was higher than with the PSPRS sum score. The Omnibus-dom test has a large power in scenarios where the effect size is homogeneous within one or more domains, but lower power than the other multiple testing procedures if the effect is in a single item only.

When simulating the data using the longitudinal IRT model, for all considered effect sizes (determining the change of the slope of the latent variable), the analysis based on the IRT model performed best in terms of power, followed by the approximate IRT model based on the weighted sum of the item scores. In these settings, also the original PSPRS sum score, the OLS test  and the Omnibus\textunderscore dom test showed a good performance. Note that in the IRT model, the treatment effect is modelled directly for the latent variable which corresponds to a treatment effect  in all items.

\paragraph{Original scoring vs. rescoring} For most of the tests we observe that the FDA scoring causes a drop in the simulated power, specifically in scenarios where there is a homogeneous effect in all items (Scenarios $\dv_{1}^\prime$, $\dv_{2}^\prime$ and $\dv_{3}^\prime$). This might be due to a loss of information by collapsing item levels. For the analysis method based on the IRT models, however, there are also scenarios where the treatment effect is in single domains only, where the power appears to increase after re-scoring. In the simulations based on the IRT model there is a tendency for a decrease in power with the rescored data.

\begin{figure}[ht!]
\centering

\subfloat[Discretised multivariate normal distribution]{\includegraphics[scale=0.55]{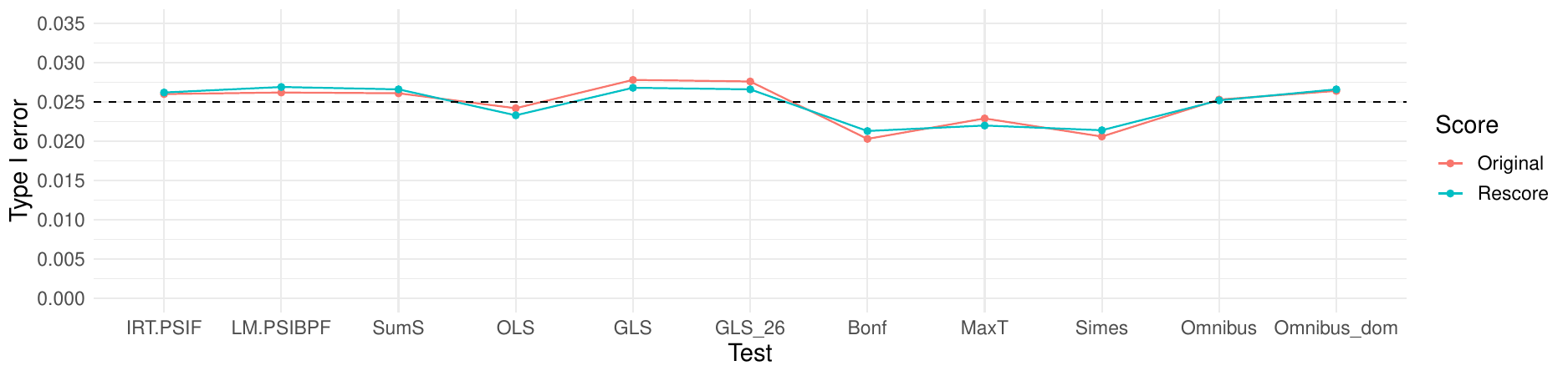}\label{Subfigure: Alphasim,DISC,LMfit,5new-tests}} \\
\subfloat[Bootstrap (without replacement)]{\includegraphics[scale=0.55]{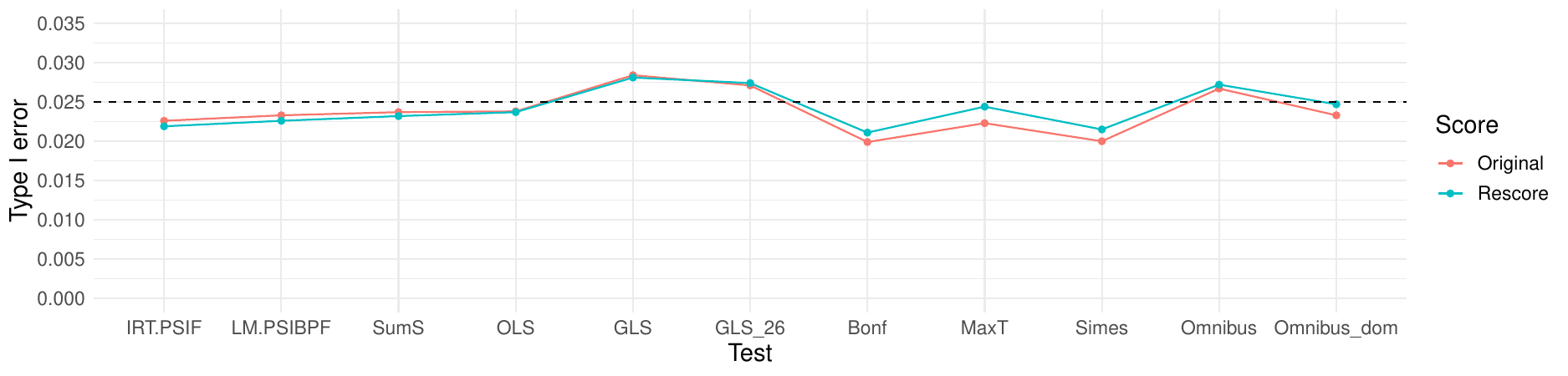}\label{Subfigure: Alphasim,BOOT,LMfit,5new-tests}} \\
\subfloat[Item scores generated from the IRT model]{\includegraphics[scale=0.55]{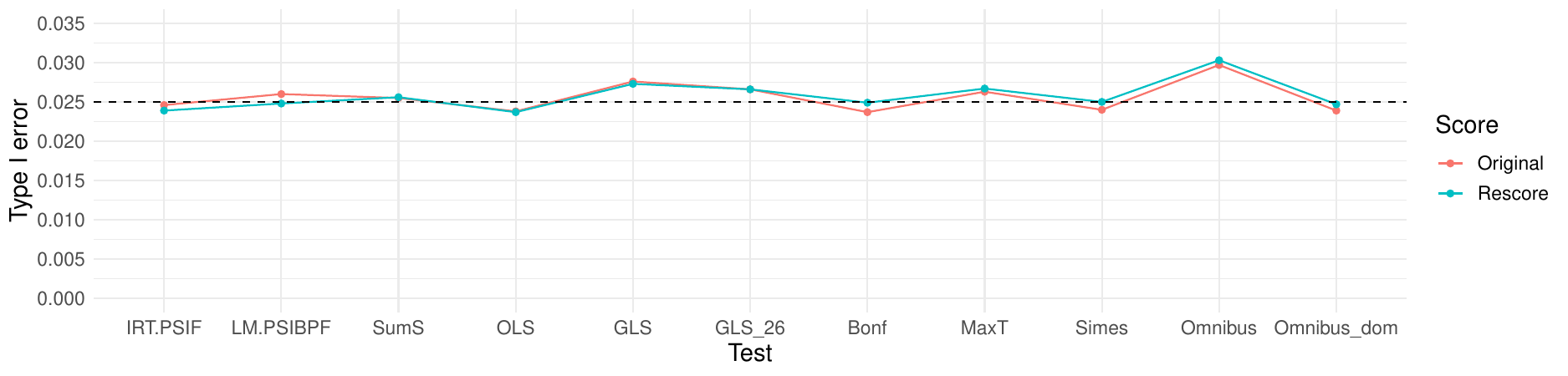}\label{Subfigure: Alphasim,IRT,LMfit,5new-tests}}

\caption{\small Type I error rates of the hypothesis tests for the different simulation approaches. The dashed lines represent the nominal significance level of 0.025 (one-sided). With 10 000 simulation runs, standard error of the estimated type I error rates (when the actual type I error is 0.025) is 0.00156. Thus, none of the methods shows an apparent inflation of the type I error rate.}
\label{Figure: Alphasim,ALLSimulationApp,LMfit,5new-tests}
\end{figure}


\begin{figure}[htp]
\centering
\vspace{-0.5cm}
\subfloat[Discretised item scores generated from multivariate normal distribution]{\includegraphics[scale=0.46]{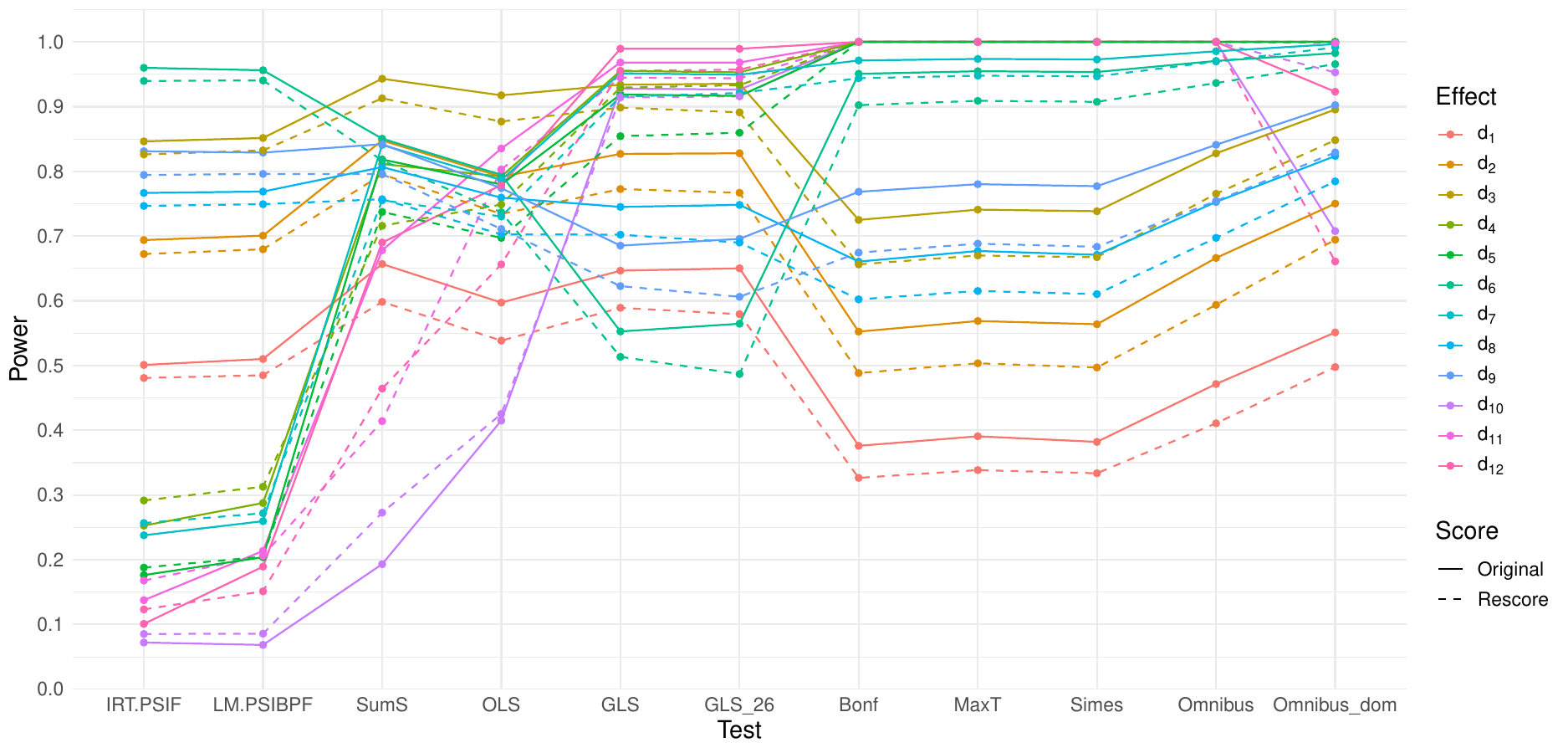}\label{Subfigure: Powersim,DISC,LMfit,5new-tests}} \\
\subfloat[Bootstrapped item scores (without replacement)]{\includegraphics[scale=0.46]{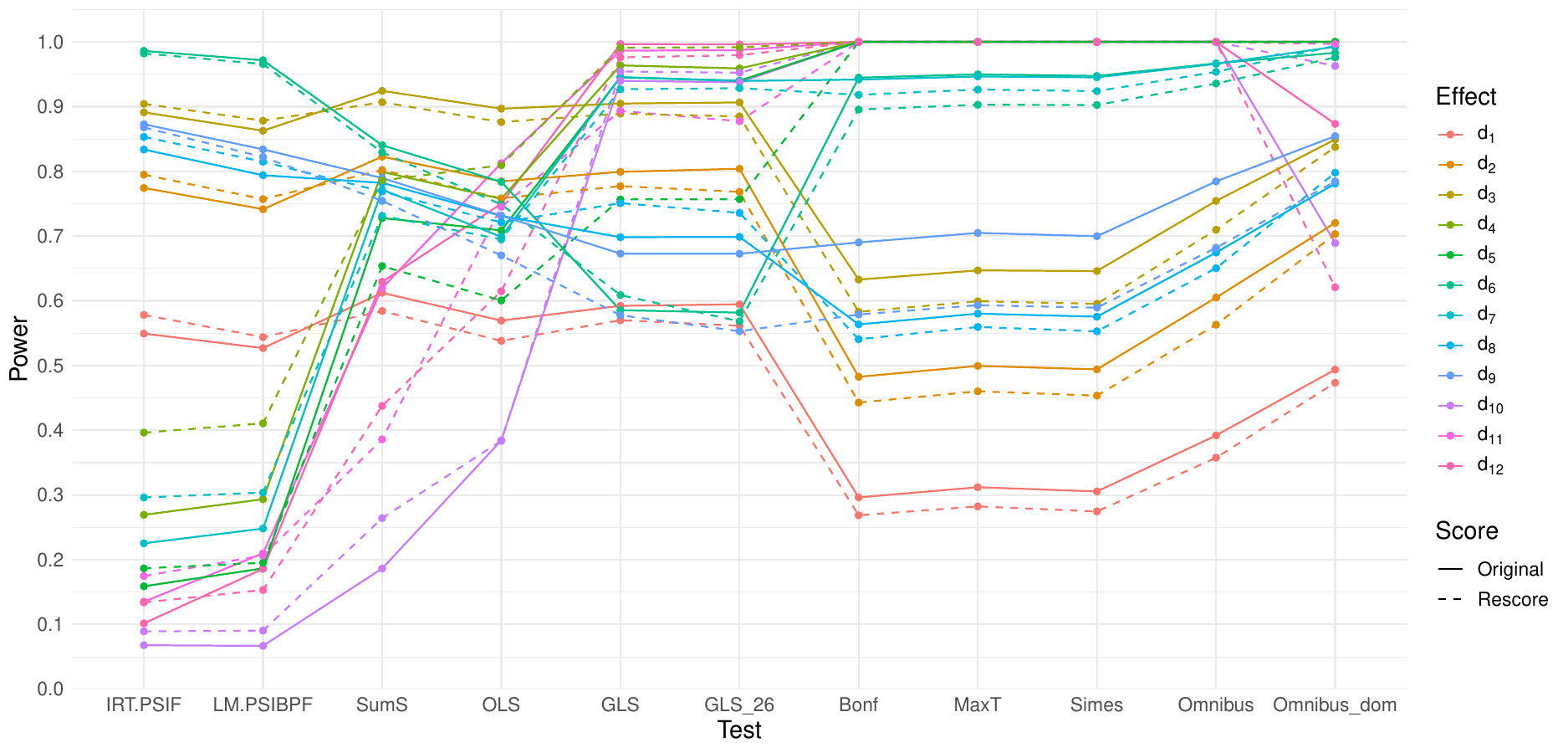}\label{Subfigure: Powersim,BOOT,LMfit,5new-tests}} \\
\subfloat[Item scores generated from the IRT model fit]{\includegraphics[scale=0.48]{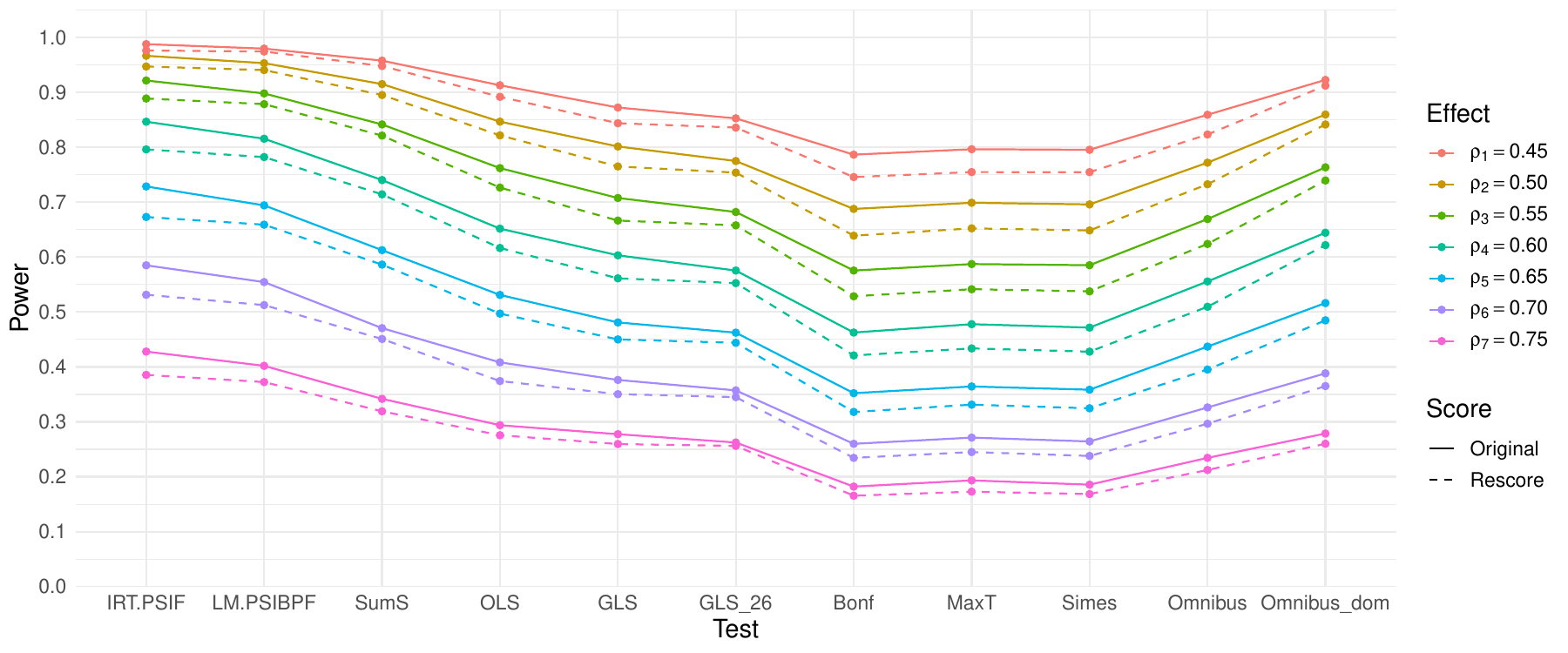}\label{Subfigure: Powersim,IRT,LMfit,5new-tests}}
\vspace{-0.5cm}
\caption{\small Power of the considered testing procedures for all simulation approaches and effect size scenarios.}
\label{Figure: Powersim,ALLSimulationApp,LMfit,5new-tests}
\end{figure}


\begin{figure}[ht!]
\begin{center}
\includegraphics[width=11cm,height=7cm]{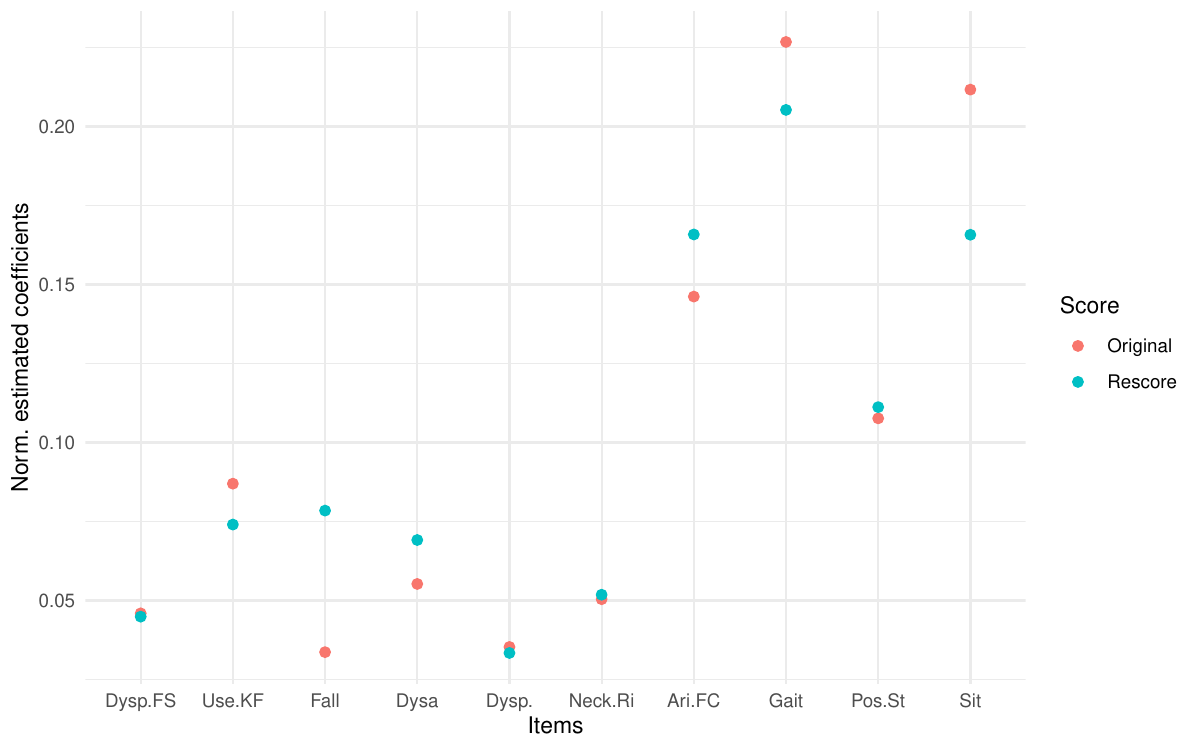} 
\end{center}
\caption{\small Linear model fit of the latent variable. The plot shows the normalised weights (summing to one) for the items with the original (red) and the FDA-recommended rescoring (blue).}\label{Figure: LM.fit.ItemsBOTH}
\end{figure}
\begin{figure}

\centering
   \subfloat{\includegraphics[scale=0.45]{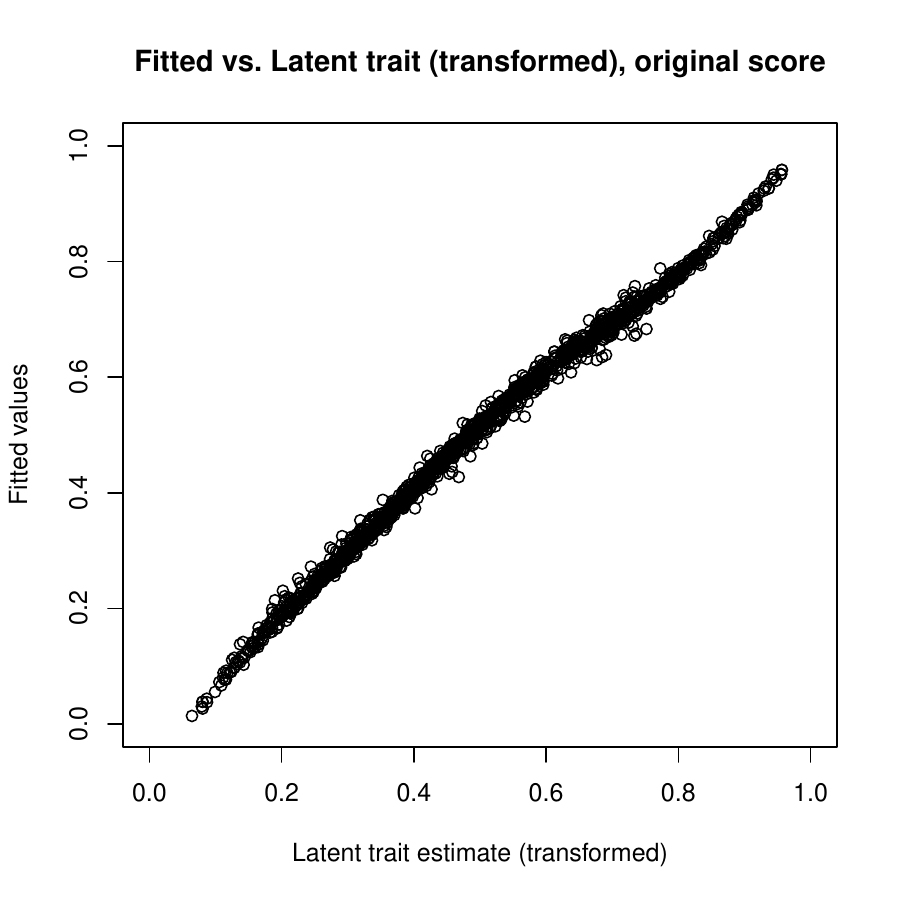}\label{fittlatentpfpullORIG}}
   \subfloat{\includegraphics[scale=0.45]{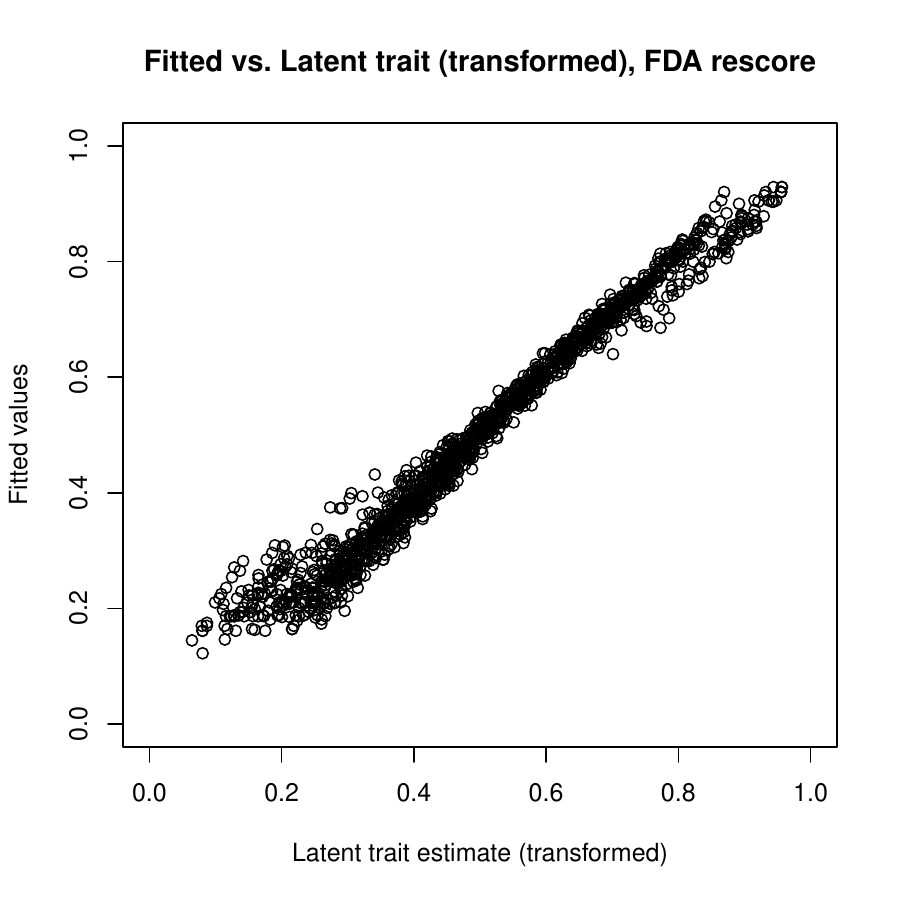}\label{fittlatentpfpullRELV}}
   \caption{Scatterplot of  the transformed estimated latent traits from the IRT model fit and the fitted values from the linear model of items.}
   \label{Figure:COR_latent_lineaModel_BOTH}
\end{figure}

\section{Reanalysis of the ABBV-8E12  trial} \label{sec-Illustration}
To illustrate the application of the different analysis methods, we re-analyse the ABBV-8E12  trial with the analysis methods considered above, we compare each dose (tilavonemab 2000, tilavonemab 4000) to placebo, in separate analyses. We included $66$, $71$, and $64$ patients in the tilavonemab 2000, tilavonemab 4000, placebo groups respectively, for which both the baseline and the Week 52 data were available for all the FDA subset of items. Note that in this data set there is a substantial proportion of missing values of item scores in later visits, specifically in Week 52, compared to baseline. This is due to the fact that the whole study was terminated early as no differences between the dose groups compared to placebo were observed in an interim analysis.
Table \ref{table.Descrp.Stat.Orig.complete} shows the descriptive statistics of the baseline and Week 52 item scores as well as the treatment effect estimates and corresponding p-values from marginal ANCOVA models with the Week 52 scores as dependent variable and the baseline score and treatment as independent variables.

Results of the different hypotheses tests defined in Section \ref{sec-Analysis.methods} are given in Table \ref{table.CombinedAnalysisTests13,23}. While, in agreement to the original analysis, none of them indicates a significant difference between groups, we find that the testing procedures yield a wide range of p-values. This reflects the fact, that the different tests focus on different aspects of deviations from the null hypothesis.

To assess the latent variable estimates obtained by a weighted sum of the item scores used in the approximate IRT-based test (Section \ref{sec-Analysis.methods}) we visually inspected the model fit, both, for the linear model fitted to the original and the rescored items.
Especially, we plot the latent variable estimate based on the linear model fit against the estimates obtained with the EAP method  in the ABBV-8E12 data set, pooling the data from all visits and treatment groups as described above (see Figure \ref{Figure:COR_latent_lineaModel_BOTH}). The figure shows that, when computing the latent variables based on the original item scores, we observe a nearly perfect model fit with minimal errors. Estimating the latent variable based on a weighted sum of the rescored items yields a somewhat larger error, especially for small values of the latent trait. The weights of the items in the resulting linear models are depicted in Figure \ref{Figure: LM.fit.ItemsBOTH} with the largest weights allocated to the items 'arising from chair', 'gait', 'postural stability' and 'sitting down' from the gait/midline exam domain.
 
\begin{figure}

\centering
   \subfloat[Tilavonemab 2000 vs placebo]{\label{Figure: weightsOLS-GLStests_Abbvie13_ORIG}\includegraphics[scale=0.45]{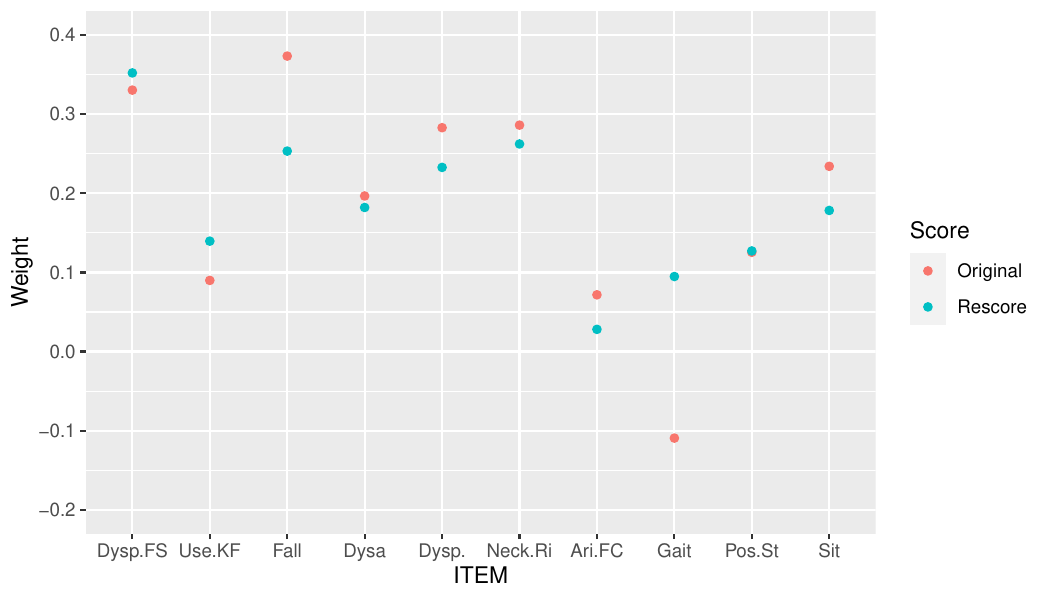}}
   \subfloat[{Tilavonemab 4000 vs placebo}]{\label{Figure: weightsOLS-GLStests_Abbvie23_ORIG}\includegraphics[scale=0.45]{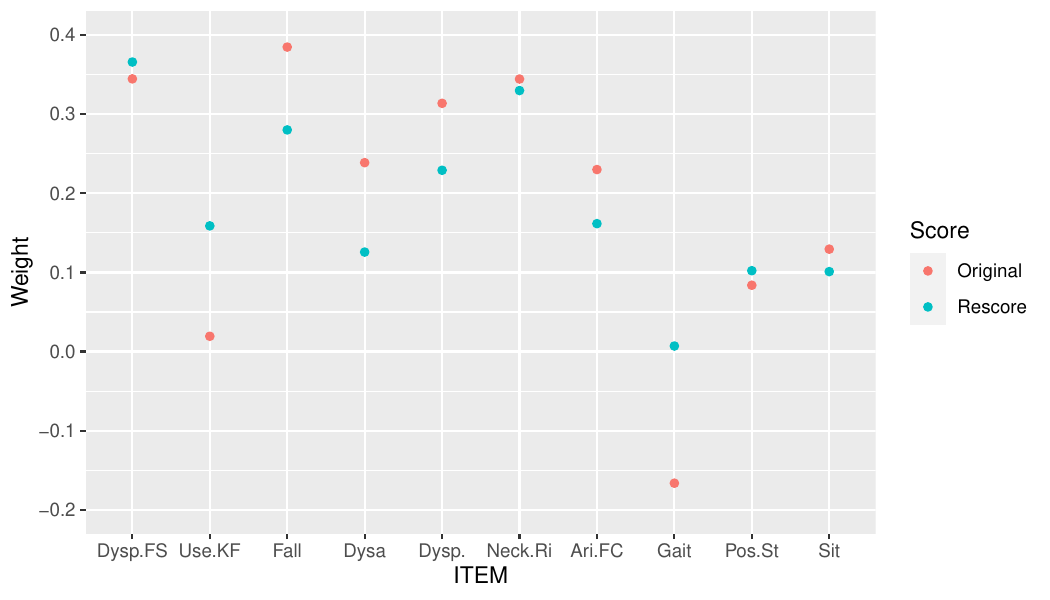}}
  \caption{Weights of the individual test statistics in the GLS test in the re-analysis of the ABBV-8E12 trial.}
\end{figure}

 The  weights for the individual t-statistics in the  GLS test are given in Figure 
 \ref{Figure: weightsOLS-GLStests_Abbvie13_ORIG}. 
For the original scores, the weight of item 26 (Gait) becomes negative. Using the FDA rescoring, we observed no negative weights. 

Note that in this illustrative example we estimated the IRT model from the same clinical trial for which the model is applied to compute the endpoint. In principle, the resulting dependence of the model estimate with the clinical trial data  may introduce a bias. Even though it is expected that this bias is not substantial, as the fit of the IRT model  does not use treatment label information, estimating the IRT model from an independent data set, avoids this issue.

\begin{table}[htp]
\linespread{1}
	\footnotesize \caption{Descriptive statistics of the 10 items (with the original scores) and marginal ANCOVA tests for comparing the respective group (2nd column) against placebo. For the baseline and Week 52 measurements, means and standard errors are reported. The one to the last column shows the treatment effect estimates (with standard error) from the marginal ANCOVA models for each of the items. Negative values correspond to a beneficial effect of treatment.}
	\label{table.Descrp.Stat.Orig.complete}
	\begin{footnotesize}
		\hfill{\renewcommand{\arraystretch}{1.4}
			\begin{center}
				\begin{tabular}{lllcccc}
					\hline 
					&&\text{Baseline}&  \text{Week 52}& \text{Difference}&  \text{ANCOVA}&\text{P-value}\\
					\hline
					\text{Dysp.FS}&\text{Til. 2000 mg}&$  0.652(0.073)$&$  0.864(0.089)$&$0.212(0.082)$&$0.06(0.104)$&$0.716$\\
					&\text{Til. 4000 mg}&$  0.662(0.078)$&$  1.028(0.112)$&$0.366(0.088)$&$0.207(0.114)$&$0.964$\\    
					&\text{Placebo}&$       0.594(0.076)$&$  0.766(0.088)$&$0.172(0.072)$&$-$&$-   $\\      
					\text{Use.KF}&\text{Til. 2000 mg}&$  1.591(0.108)$&$  2.242(0.133)$&$0.652(0.119)$&$0.001(0.167)$&$0.503$\\    
					&\text{Til. 4000 mg}&$  1.577(0.098)$&$  2.296(0.116)$&$0.718(0.105)$&$0.062(0.156)$&$0.653$\\    
					&\text{Placebo}&$       1.578(0.106)$&$  2.234(0.127)$&$0.656(0.138)$&$-$&$-   $\\ 
					\text{Fall}&\text{Til. 2000 mg}&$  2.409(0.126)$&$  2.712(0.155)$&$0.303(0.146)$&$-0.149(0.194)$&$0.221$\\    
					&\text{Til. 4000 mg}&$  2.056(0.122)$&$  2.451(0.151)$&$0.394(0.137)$&$-0.229(0.189)$&$0.114$\\    
					&\text{Placebo}&$   2.219(0.147)$&$  2.766(0.149)$&$0.547(0.157)$&$-$&$-   $\\     
					\text{Dysa.}&\text{Til. 2000 mg}&$  1.50 (0.104)$&$  2.03(0.124) $&$0.53(0.099) $&$0.186(0.133) $&$0.918$\\   
					&\text{Til. 4000 mg}&$ 1.563(0.091)$&$  2.042(0.111)$&$0.479(0.098)$&$0.152(0.13)$&$0.878$\\
					&\text{Placebo}&$   1.641(0.098)$&$  1.938(0.104)$&$0.297(0.099)$&$-$&$-   $\\      
					\text{Dysp.}&\text{Til. 2000 mg}&$  1.197(0.117)$&$  1.439(0.137)$&$0.242(0.15) $&$-0.171(0.189)$&$0.183$\\    
					&\text{Til. 4000 mg}&$  1.028(0.104)$&$  1.521(0.137)$&$0.493(0.131)$&$-0.016(0.184) $&$0.466$\\   
					&\text{Placebo}&$   1.047(0.123)$&$  1.547(0.148)$&$0.50 (0.149)$&$-$&$-   $\\      
					\text{Neck.Ri}&\text{Til. 2000 mg}&$  1.773(0.101)$&$  2.182(0.128)$&$0.409(0.108)$&$-0.011(0.148)$&$0.471$\\    
					&\text{Til. 4000 mg}&$  1.648(0.108)$&$  2.028(0.125)$&$0.38(0.097) $&$-0.07(0.139) $&$0.308$\\   
					&\text{Placebo}&$   1.562(0.111)$&$  2.031(0.135)$&$0.469(0.104)$&$-$&$-   $\\       
					\text{Ari.FC}&\text{Til. 2000 mg}&$  2.076(0.159)$&$  2.879(0.158)$&$0.803(0.163)$&$-0.035(0.194)$&$0.429$\\    
					&\text{Til. 4000 mg}&$  2.183(0.145)$&$  2.873(0.149)$&$0.69(0.123) $&$-0.111(0.174)$&$0.262$\\    
					&\text{Placebo}&$   2.062(0.157)$&$  2.906(0.165)$&$0.844(0.145)$&$-$&$-   $\\      
					\text{Gait}&\text{Til. 2000 mg}&$  1.985(0.117)$&$  2.606(0.114)$&$0.621(0.094)$&$0.008(0.126)$&$0.527$\\    
					&\text{Til. 4000 mg}&$  1.944(0.10)$&$   2.451(0.106)$&$0.507(0.087)$&$-0.121(0.123)$&$0.163$\\   
					&\text{Placebo}&$  1.812(0.111)$&$  2.484(0.118)$&$0.672(0.10) $&$-$&$-   $\\      
					\text{Pos.St}&\text{Til. 2000 mg}&$  2.242(0.13)$&$   2.879(0.132)$&$0.636(0.114)$&$-0.071(0.158)$&$0.327$\\    
					&\text{Til. 4000 mg}&$  2.056(0.139)$&$  2.761(0.141)$&$0.704(0.114)$&$-0.079(0.159)$&$0.311$\\   
					&\text{Placebo}&$   2.141(0.141)$&$  2.891(0.139)$&$0.75 (0.135)$&$-$&$-   $\\       
					\text{Sit}&\text{Til. 2000 mg}&$ 1.712(0.107)$&$  2.394(0.143)$&$0.682(0.102)$&$-0.166(0.155)$&$0.143$\\    
					&\text{Til. 4000 mg}&$ 1.718(0.101)$&$  2.437(0.133)$&$0.718(0.121)$&$-0.122(0.162)$&$0.227$\\    
					&\text{Placebo}&$  1.672(0.126)$&$  2.531(0.13) $&$0.859(0.126)$&$-$&$-   $\\  
					\hline
				\end{tabular}
			\end{center}
		}
	\end{footnotesize}
 \linespread{1}
\end{table}

\begin{table}[htp]
\linespread{1}
	\caption{Descriptive statistics of the 10 items (with the FDA re-scores) and marginal ANCOVA tests for comparing the respective group (2nd column) against placebo. Similar to Table \ref{table.Descrp.Stat.Orig.complete}, for the baseline and Week 52 measurements, means and standard errors are reported. The one to the last column shows the treatment effect estimates (with standard error) from the marginal ANCOVA models for each of the items. Negative values correspond to a beneficial effect of treatment.}
	\label{table.Descrp.Stat.RESC.complete}
	\begin{footnotesize}
		\hfill{\renewcommand{\arraystretch}{1.4}
			\begin{center}
				\begin{tabular}{lllcccc}
					\hline 
					&&\text{Baseline}&  \text{Week 52}& \text{Difference}&  \text{ANCOVA} &\text{P-value}\\
					\hline
					\text{Dysp.FS}&\text{Til. 2000 mg}&$  0.652(0.073)$&$       0.864(0.089)$&$                0.212(0.082)$&$0.06(0.104)$&$0.716$\\    
					&\text{Til. 4000 mg}&$ 0.662(0.078)$&$       1.028(0.112)$&$               0.366(0.088)$&$0.207(0.114)$&$0.964$\\    
					&\text{Placebo}&$  0.594(0.076)$&$       0.766(0.088)$&$                0.172(0.072)$&$-$&$-   $\\       
					\text{Use.KF}&\text{Til. 2000 mg}&$  1.561(0.099)$&$       2.106(0.11)$&$                 0.545(0.106)$&$-0.012(0.143)$&$0.467$\\   
					&\text{Til. 4000 mg}&$ 1.563(0.094)$&$       2.211(0.102)$&$                0.648(0.092)$&$0.093(0.135)$&$0.754$\\    
					&\text{Placebo}&$  1.578(0.106)$&$       2.125(0.108)$&$                0.547(0.128)$&$-$&$-   $\\       
					\text{Fall}&\text{Til. 2000 mg}&$  1.167(0.051)$&$       1.348(0.07)$&$                 0.182(0.064)$&$-0.016(0.092)$&$0.432$\\    
					&\text{Til. 4000 mg}&$ 1.042(0.047)$&$       1.268(0.063)$&$                0.225(0.061)$&$-0.037(0.088)$&$0.339$\\    
					&\text{Placebo}&$  1.094(0.062)$&$       1.328(0.071)$&$                0.234(0.076)$&$-$&$-   $\\       
					\text{Dysa.}&\text{Til. 2000 mg}&$  1.015(0.055)$&$       1.303(0.065)$&$                0.288(0.071)$&$0.09(0.082)$&$0.865$\\    
					&\text{Til. 4000 mg}&$ 1.056(0.049)$&$       1.225(0.061)$&$                0.169(0.063)$&$-0.004(0.079)$&$0.482$\\    
					&\text{Placebo}&$  1.031(0.054)$&$       1.219(0.057)$&$                0.188(0.062)$&$-$&$-   $\\\       
					\text{Dysp.}&\text{Til. 2000 mg}&$  1.197(0.117)$&$       1.439(0.137)$&$                0.242(0.15)$&$-0.171(0.189)$&$0.183$\\    
					&\text{Til. 4000 mg}&$ 1.028(0.104)$&$       1.521(0.137)$&$                0.493(0.131)$&$-0.016(0.184)$&$0.466$\\    
					&\text{Placebo}&$  1.047(0.123)$&$       1.547(0.148)$&$                0.5(0.149)$&$-$&$-   $\\       
					\text{Neck.Ri}&\text{Til. 2000 mg}&$  1.091(0.060)$&$       1.455(0.092)$&$                0.364(0.077)$&$0.089(0.112)$&$0.786$\\    
					&\text{Til. 4000 mg}&$ 1.127(0.066)$&$       1.338(0.085)$&$                0.211(0.072)$&$-0.05(0.107)$&$0.32 $\\   
					&\text{Placebo}&$  1.062(0.066)$&$       1.344(0.092)$&$                0.281(0.085)$&$-$&$-   $\\       
					\text{Ari.FC}&\text{Til. 2000 mg}&$  0.606(0.094)$&$       1.152(0.102)$&$                0.545(0.097)$&$-0.075(0.127)$&$0.277$\\    
					&\text{Til. 4000 mg}&$ 0.634(0.088)$&$       1.127(0.10)$&$                  0.493(0.085)$&$-0.12(0.122)$&$0.163$\\   
					&\text{Placebo}&$   0.562(0.091)$&$       1.203(0.102)$&$                0.641(0.101)$&$-$&$-   $\\       
					\text{Gait}&\text{Til. 2000 mg}&$  1(0.101)$&$           1.455(0.092)$&$                0.455(0.081)$&$-0.068(0.11)$&$0.269$\\    
					&\text{Til. 4000 mg}&$ 0.958(0.097)$&$       1.338(0.087)$&$                0.38(0.076)$&$-0.16(0.106)$&$0.066$\\    
					&\text{Placebo}&$  0.828(0.101)$&$       1.422(0.102)$&$                0.594(0.094)$&$-$&$-   $\\      
					\text{Pos.St}&\text{Til. 2000 mg}&$  1.303(0.116)$&$       1.894(0.128)$&$                0.591(0.108)$&$-0.052(0.15)$&$0.364$\\    
					&\text{Til. 4000 mg}&$ 1.155(0.121)$&$       1.817(0.127)$&$                0.662(0.108)$&$-0.034(0.149)$&$0.409$\\     
					&\text{Placebo}&$  1.266(0.114)$&$       1.922(0.13)$&$                 0.656(0.118)$&$-$&$-   $\\       
					\text{Sit}&\text{Til. 2000 mg}&$  0.788(0.093)$&$       1.455(0.13)$&$                 0.667(0.097)$&$-0.102(0.15)$&$0.249$\\    
					&\text{Til. 4000 mg}&$ 0.761(0.093)$&$       1.493(0.12)$&$                 0.732(0.108)$&$-0.048(0.152)$&$0.375$\\    
					&\text{Placebo}&$  0.797(0.105)$&$       1.562(0.122)$&$                0.766(0.121)$&$-$&$-   $\\   
					\hline
				\end{tabular}
			\end{center}
		}
	\end{footnotesize}
 \linespread{1}
\end{table}

\begin{table}[htp]
  \centering
	\subfloat{
		\resizebox{\textwidth}{!}{
			\begin{tabular}{llccccccccccc}
				\hline
			&&\text{IRT-PSIF}&\text{LM-PSIBPF}&\text{SumS}&\text{OLS}&\text{GLS}&\text{GLS-26}&\text{Bonf}&\text{MaxT}&\text{Simes}&\text{Omnibus}&\text{Omnibus-dom}\\
				\hline
			\text{Original score}&\text{Til. 2000 mg vs Placebo}&$ 0.30$&$ 0.34$&$ 0.23$&$ 0.40$&$ 0.34$&$ 0.37$&$ 1.00$&$ 0.64$&$ 0.79$&$ 0.73$&$ 0.52$\\
			&\text{Til. 4000 mg vs Placebo}&$ 0.29$&$ 0.31$&$ 0.35$&$ 0.41$&$ 0.55$&$ 0.53$&$ 1.00$&$ 0.56$&$ 0.72$&$ 0.64$&$ 0.42$\\
			\hline
			\text{FDA re-score}&\text{Til. 2000 mg vs Placebo}&$ 0.13$&$ 0.18$&$ 0.24$&$ 0.44$&$ 0.54$&$ 0.54$&$ 1.00$&$ 0.73$&$ 0.86$&$ 0.80$&$ 0.55$\\
				&\text{Til. 4000 mg vs Placebo}&$ 0.27$&$ 0.27$&$ 0.39$&$ 0.39$&$ 0.60$&$ 0.60$&$ 0.66$&$ 0.39$&$ 0.60$&$ 0.55$&$ 0.45$\\
				\hline
			\end{tabular}
		}
		\label{subtable.AnalysisTest2000(Test13)}
	}
	\caption{One-sided p-values of the hypothesis tests in the re-analysis of the ABBV-8E12 trial using different analysis methods, based on the original scoring and the FDA re-scoring.}
	\label{table.CombinedAnalysisTests13,23}
\end{table}

\section{Discussion} \label{sec-Discussion}

In this paper, we assessed a range of testing approaches to compare treatment groups in multivariate endpoints in a simulation study in the setting of PSP. We investigated multiple data generation approaches and effect size scenarios to assess the robustness of findings. Besides classical approaches to test multivariate endpoints, we also consider tests based on estimates of the latent variables computed from an IRT model which represent the patients disease status. In addition, we consider an approximate version of the corresponding outcome variable, which is defined as a weighted average of the individual item scores and fits the IRT based estimate surprisingly well.

This study has several limitations. First, we assumed that the clinical trial has no missing values or dropouts. Dropouts can lead to lower power, due to a lower sample size, but also can introduce bias if the distribution of the observed outcomes differs from the distribution of the unobserved values which are missing. Due to repeated measurements of the PSPRS items, approaches using early outcome variables can be used to account for missing data, e.g., based on mixed models that model the course of the disease. In future work, the proposed analysis approaches can be extended to such models. \citeauthor{vickerstaff2021comparison} \cite{vickerstaff2021comparison} assessed a number of approaches to analyse multiple correlated outcomes focusing on settings with missing data, limiting the analysis, however, to the case of  two endpoints only.

We utilized simulation approaches based on a discretised multivariate normal distribution and Bootstrap, both flexible tools for simulating data across diverse effect size scenarios, including uniform and heterogeneous item-level and domain-wise effect sizes. Our third approach utilized a longitudinal IRT model, focusing on the effect on the underlying latent variable. These approaches are particularly relevant for calculating power based on effect size assumptions for specific domains or items in the 10-item version of the PSPRS. 

The simulation study demonstrated that none of the considered testing procedures is uniformly optimal but the most powerful test depends on the specific configuration of effect sizes and the data generating mechanism. 
The test based on the PSPRS sum score of the FDA recommended subset of items performed well in scenarios where there is an homogeneous effect across items. Also in settings where there is an effect in some of the domains only, it still provides moderate power. Similarly, the GLS, OLS, the Omnibus tests and the tests based on item response models have a larger power than the tests based on multiplicity adjustments of marginal tests. In treatment effect scenarios, where the treatment effect is limited to few domains or items, tests based on multiple testing procedures have a higher power. Also the Omnibus tests have a high power in these settings, however, for those type I error control under dependence can only be demonstrated by simulations.
When simulating the data based on an IRT model, the IRT based analysis appears to be the most powerful. This is not unexpected, as it is the correct model in this case. 
Another observation is that, for most tests and scenarios, the FDA recommended scoring of the items causes a reduction in the simulated power. 

Because all tests, with the exception of the IRT based test, test the strong null hypothesis as defined in Section \ref{sec-Analysis.methods}, they can be extended to multiple tests using the closed testing principle. Such multiple tests test for each item the individual null hypotheses that the expected score is large or equal under treatment than under control controlling the FWER in the strong sense. Rejection of an individual null hypothesis not only allows to conclude that there is an effect in any item (as follows from rejection of the strong null hypothesis), but also to identify in which item the effect is. Tests based on the original PSPRS or the approximate IRT-based test allow for a conclusion on the (weighted) average of the item scores, which may have a clearer clinical interpretation than rejection of the strong null hypothesis. 

As the optimal testing procedure in terms of statistical power depends on the specific effect size patterns assumed, identifying plausible treatment effect patterns is crucial when planning a study. For instance, treatments with disease-modifying effects that slow disease progression are expected to impact all items, either uniformly or as modeled by the longitudinal IRT model. Conversely, symptomatic treatments are likely to influence certain items or domains more selectively. Given scenarios with assumed treatment effects, simulation studies can guide the selection of the most suitable testing strategy. When there is significant uncertainty regarding the treatment effect pattern, a maximin criterion can be applied to choose methods that offer the highest minimum power across all plausible scenarios.

In conclusion, our study underscores that there is no one-size-fits-all testing procedure for evaluating treatment effects using PSPRS items; the optimal method varies based on the specific effect size patterns. The efficiency of the PSPRS sum score, while generally robust and straightforward to apply, varies depending on the specific patterns of effect sizes encountered and more powerful alternatives are available in specific settings.



\printbibliography

\begin{description}

\item[Acknowledgements]

~\\This publication is based on research using data from data contributors AbbVie that has been made available through Vivli, Inc. Vivli has not contributed to or approved, and is not in any way responsible for, the contents of this publication. The authors thank Amylyx Pharmaceuticals for providing the PSPRS rescoring guideline, based on an FDA PreIND meeting.
The authors used ChatGPT, OpenAI’s large-scale language-generation model, for English language editing of part of the text. 

\item [Author contributions]
~\\F.K., G.H. and M.P. designed the study. G.H. and F.H. established data access and harmonized data across datasets.  E.Y. prepared the initial draft of the text, performed the statistical analysis and simulations. M.G. and M.K. developed the IRT model to be used in the IRT model based simulation. S.Z. and R.R contributed to the implementation of the simulation study. All authors discussed the results, provided comments and reviewed the manuscript.

\item[Funding]

~\\ E.Y., M.G., F.K., G.H., M.K and M.P. received funding from the European Joint Programme on Rare Diseases (Improve-PSP). G.H. was supported by the German Federal Ministry of Education and Research (BMBF: 01KU1403A EpiPD; 01EK1605A HitTau), Deutsche Forschungsgemeinschaft (DFG, German Research Foundation) under Germany’s Excellence Strategy within the framework of the Munich Cluster for Systems Neurology (EXC 2145 SyNergy – ID 390857198), DFG grants (HO2402/6-2, HO2402/18-1 MSAomics), the German Federal Ministry of Education and Research (BMBF, 01KU1403A EpiPD; 01EK1605A HitTau; 01DH18025 TauTherapy; CurePML EN2021-039); Niedersächsisches Ministerium für Wissenschaft und Kunst (MWK)/VolkswagenStiftung (Niedersächsisches Vorab), Petermax-Müller Foundation (Etiology and Therapy of Synucleinopathies and Tauopathies).


\item[Competing interests] 
G.H. has ongoing research collaborations with Roche, UCB, Abbvie; serves as a consultant for Abbvie, Alzprotect, Amylyx, Aprineua, Asceneuron, Bayer, Bial, Biogen, Biohaven, Epidarex, Ferrer, Kyowa Kirin, Lundbeck, Novartis, Retrotope, Roche, Sanofi, Servier, Takeda, Teva, UCB; received honoraria for scientific presentations from Abbvie, Bayer, Bial, Biogen, Bristol Myers Squibb, Kyowa Kirin, Pfizer, Roche, Teva, UCB, Zambon; holds a patent on Treatment of Synucleinopathies (US 10,918,628 B2, EP 17 787 904.6-1109 / 3 525 788); received publication royalties from Academic Press, Kohlhammer, and Thieme.

\end{description}



\section*{Supplementary material}

# Supplementary Material A

December 13, 2023

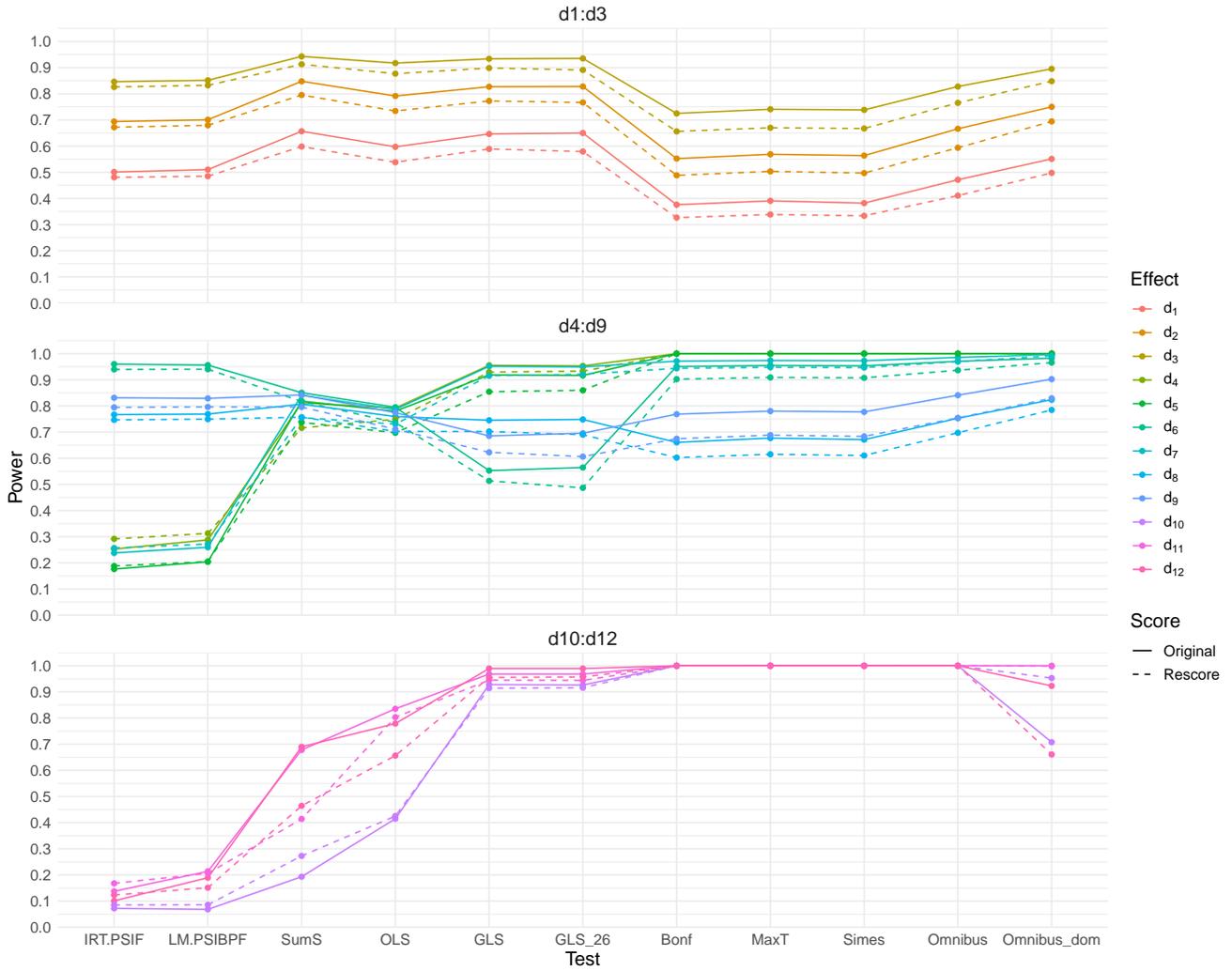

Figure 1: Power of the considered testing procedures for the simulations based on discretised multivariate normal item scores.

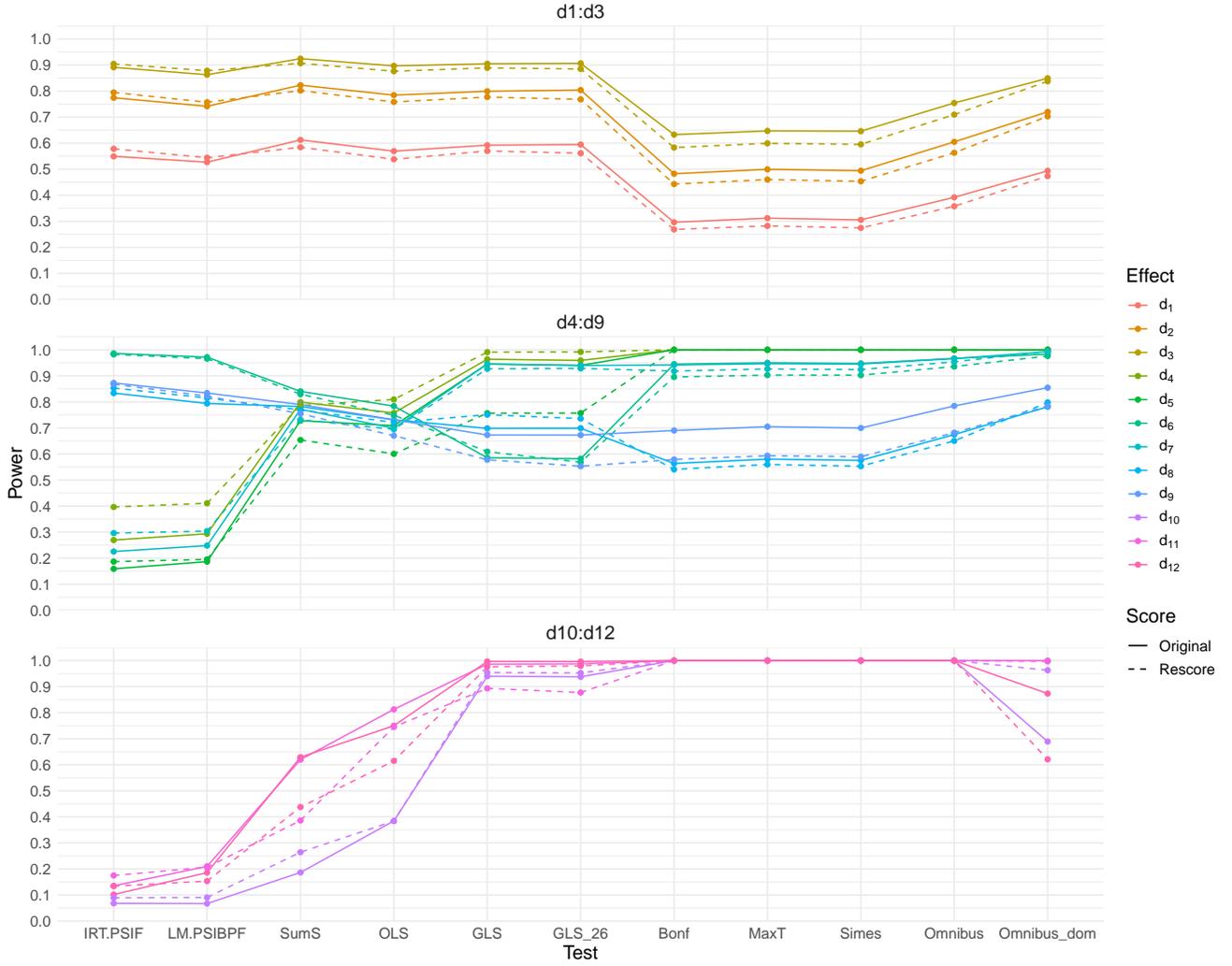

Figure 2: Power of the considered testing procedures for the simulations based on the Bootstrap method.

# 1 Multivariate tests

This section aims at providing technical details about the OLS and GLS tests. Let $n_1$ and $n_2$ denote, respectively, the number of subjects for the control and test groups. Further, $m \geq 2$ stands for the number of endpoints for each treatment group. Additionally, $x_{ijk}$ represents the measurement on the $k$th endpoint for the $j$th subject in the $i$th treatment group. For treatment group $i$, the vector of observations, $\boldsymbol{x}_{ij} = (x_{ij1}, x_{ij2}, \ldots, x_{ijm})'$, $j = 1, 2, \ldots, n_i$ has mean vector $\boldsymbol{\mu}_i = (\mu_{i1}, \mu_{i2}, \ldots, \mu_{im})'$ and covariance matrix $\boldsymbol{\Sigma}_i$. The covariance matrix $\boldsymbol{\Sigma}_i$ is based on pairwise covariances and is of dimension $m \times m$. Furthermore, $\boldsymbol{\delta} = \boldsymbol{\mu}_2 - \boldsymbol{\mu}_1 = (\delta_1, \delta_2, \ldots, \delta_m)'$ denotes the vector of mean differences between the treatment group and the control group. The interest is to test a one-sided alternative to detect a treatment effect as follows

$$\begin{cases} H_0 : \delta_k \geq 0, \forall k \in \{1, \ldots, m\} \\ H_1 : \exists\ k \in \{1, \ldots, m\} : \ \delta_k \leq 0. \end{cases}$$

This hypothesis test indicates that, under the alternative the test treatment is supposed to have a nonzero effect in at least one of the components of the multivariate endpoint.

## 1.1 O'Brien's OLS and GLS tests (homoscedastic case)

For the OLS and GLS tests we used similar ideas as presented in [1, 2]. There the t-statistics are based on the vector of mean differences. Here we estimate the t-statistics from the marginal ANCOVA models. We denote by $t_k = \hat{\beta}_k/\text{SE}(\hat{\beta}_k)$ the standardised treatment effect estimate for $k$th endpoint, in which $\hat{\beta}_k$ is the estimated coefficient of the treatment variable in the marginal ANCOVA model of kth item and $\text{SE}(\hat{\beta}_k)$ indicates the corresponding standard error estimate. The vector of t-statistics is then assumed (under the null) to have a multivariate normal distribution with the zero mean and the correlation matrix $\boldsymbol{R}$ of the treatment coefficients from the marginal ANCOVA models. Assuming that there is a consistent estimate of $\boldsymbol{R}$, O'Brien [1] proposed the OLS test statistic as

$$t_{OLS} = \frac{\boldsymbol{j}'\boldsymbol{t}}{\sqrt{\boldsymbol{j}'\hat{\boldsymbol{R}}\boldsymbol{j}}}. \tag{1}$$

Here $\boldsymbol{j}$ represents a vector of 1's with an appropriate dimension. Additionally, $\boldsymbol{t}$ is the vector of t-statistics, with each component denoted by $t_k$, described above.

As mentioned in earlier sections, we considered the available data from the ABBV-8E12 study for each of the 10 endpoints at two time points: baseline and week 52. To derive the vector of individual t-statistics in the OLS test statistic (vector of $\boldsymbol{t}$ on the right-hand side of Eq. 1), we fitted multiple marginal models to data for each endpoint. Finally, we calculated Eq. 1 using the estimated correlation matrix $\hat{\boldsymbol{R}}$ of the treatment coefficients from the marginal ANCOVA models, $\hat{\boldsymbol{R}}$. This calculation was performed using the `multcomp` package in R. Each $t_k$ is marginally $t$-distributed (under $H_0$) with $n_1 + n_2 - 3$ degrees of freedom, where the degrees of freedom are determined by including two covariates (baseline and treatment) in the fitted marginal ANCOVA models.

Due to correlation between the endpoints in each group, one may prefer a test statistic based on the generalised least squares (GLS) estimate of t-statistics. In this case, the corresponding GLS test statistic is represented as [1]

$$t_{GLS} = \frac{\boldsymbol{j}'\hat{\boldsymbol{R}}^{-1}\boldsymbol{t}}{\sqrt{\boldsymbol{j}'\hat{\boldsymbol{R}}^{-1}\boldsymbol{j}}}. \tag{2}$$

Both the OLS and GLS test statistics are weighted sums of individual test statistics for $m$ endpoints. The former uses equal weights, while the latter employs unequal weights determined by $\hat{\boldsymbol{R}}^{-1}$. For small samples, the OLS test statistic is approximated by the t-distribution with $n_1 + n_2 - 2m$ d.f. We have employed the modification, presented in [2], as mentioned in Subsection 3.2 using $0.5(n1 + n2 - 3)(1 + 1/m^2)$.

## 2 IRT model estimation

An item response model describes the relationship between the underlying latent trait and the probability of a certain response (score) for a specific item. The model assumes that the questionnaire/data in hand measures a latent variable, and depending on the number of parameters used in the model, the assumption is that the latent variable prediction is dependent on those parameters and is estimated simultaneously along with those parameters using an estimation algorithm. We have considered a two parameter GR model, which is usually used to model graded items. It assumes that the item difficulty and its discrimination capacity directly affects the probability of a certain score, which in turn is linked to a certain latent variable. The probability for a subject $i$ to have at least a score of $s$ for item $j$ is [3]:

$$P(y_{ij} \geq s) = \frac{1}{1 + e^{a_j(b_{j,s} - \psi_i)}}, \quad (3)$$

where $a_j$ denotes the discrimination parameter and $b_{j,s}$ is the difficulty parameter for the s-th step of the item. Besides $\psi_i$ is the representation of the unobserved latent trait. The discrimination parameter $a$ quantifies how well an item differentiates between individuals with different levels of the latent trait being measured. The difficulty parameters $b_1 - b_4$ indicate the level of the latent trait at which an individual is equally likely to choose one response category over another. The probability to have a certain score $s$ is calculated as

$$P(y_{ij} = s) = P(y_{ij} \geq s) - P(y_{ij} \geq s+1).$$

Additionally $P(y_{ij} \geq 0) = 1$ and $P(y_{ij} \geq S_j + 1) = 0$ hold.

Using the aggregated data at the baseline and follow up visits from the the FDA-recommended 10-item version of the PSPRS from the ABBV-8E12 trial, the estimated parameters for the GR model, Eq. (3), are presented in the following table.

Table 1: Estimation of the IRT model parameters based on the ABBV-8E12 trial dataset, where the aggregated data from all treatment groups and visits are used to fit the IRT model.

|         | Original score | | | | | FDA score | | | | |
|---------|-------|--------|--------|--------|-------|-------|--------|--------|-------|-------|
|         | $a$   | $b_1$  | $b_2$  | $b_3$  | $b_4$ | $a$   | $b_1$  | $b_2$  | $b_3$ | $b_4$ |
| Dysp.FS | 0.918 | −0.558 | 2.581  | 4.439  | 6.555 | 0.918 | −0.554 | 2.586  | 4.436 | 6.537 |
| Use.KF  | 1.673 | −2.126 | −0.332 | 0.752  | 2.423 | 1.677 | −2.132 | −0.328 | 0.750 | −     |
| Fall    | 0.919 | −3.510 | −1.491 | 0.239  | 1.453 | 1.197 | −2.894 | 1.243  | −     | −     |
| Dysa.   | 1.118 | −2.806 | −0.400 | 1.765  | 3.524 | 1.152 | −2.765 | 1.759  | −     | −     |
| Dysp.   | 0.942 | −1.000 | 0.561  | 1.969  | 5.845 | 0.929 | −1.009 | 0.571  | 1.994 | 5.905 |
| Neck.Ri | 0.967 | −2.740 | −0.701 | 1.234  | 3.437 | 1.013 | −2.656 | 1.204  | 3.314 | −     |
| Ari.FC  | 3.370 | −1.439 | −0.541 | −0.253 | 0.674 | 3.315 | −0.244 | 0.682  | −     | −     |
| Gait    | 3.772 | −2.051 | −0.669 | 0.131  | 1.854 | 4.132 | −0.648 | 0.132  | −     | −     |
| Pos.St  | 2.429 | −1.685 | −0.940 | −0.185 | 0.839 | 2.524 | −0.920 | −0.176 | 0.831 | −     |
| Sit     | 3.420 | −1.558 | −0.439 | 0.424  | 1.577 | 3.316 | −0.434 | 0.432  | 1.591 | −     |

# SUPPLEMENTARY MATERIAL B

| ORIGINAL | recommended change | FDA-recommended 10-item version |
|---|---|---|
| **I. HISTORY (from patient or other informant)** | | |
| 1. Withdrawal (from patient or other informant) | eliminate | |
|     0 None | | |
|     1 Follows conversation in a group, may respond spontaneously, but rarely if ever initiates exchanges | | |
|     2 Rarely or never follows conversation in a group | | |
| 2. Irritability (from patient or other informant) | eliminate | |
|     0 No increase in aggressiveness | | |
|     1 Increased, but not interfering with family interactions | | |
|     2 Interfering with family interactions | | |
| 3. Dysphagia for solids (from patient or other informant) | | 3. Dysphagia for solids (from patient or other informant) |
|     0 Normal; no difficulty with full range of food textures | |     0 Normal; no difficulty with full range of food textures |
|     1 Tough foods must be cut up into small pieces | |     1 Tough foods must be cut up into small pieces |
|     2 Requires soft solid diet | |     2 Requires soft solid diet |
|     3 Requires pureed or liquid diet | |     3 Requires pureed or liquid diet |
|     4 Tube feeding required for some or all feeding | |     4 Tube feeding required for some or all feeding |
| 4. Using knife and fork, buttoning clothes, washing hands and face (rate the worst) | | 4. Using knife and fork, buttoning clothes, washing hands and face (rate the worst) |
|     0 Normal | |     0 Normal |
|     1 Somewhat slow but no help required | |     1 Somewhat slow but no help required |
|     2 Extremely slow; or occasional help needed | |     2 Extremely slow; or occasional help needed |
|     3 Considerable help needed but can do some things alone | collapse old 3 and 4 into new 3 |     3 Considerable help needed |
|     4 Requires total assistance | | |
| 5. Falls (average frequency if patient attempted to walk unaided) | | 5. Falls (average frequency if patient attempted to walk unaided) |
|     0 None in the past year | |     0 None in the past year |
|     1 < 1 per month; gait may otherwise be normal | collapse old 1, 2, and 3 into new 1 |     1 <= 30 per month |
|     2 1-4 per month | | |
|     3 5-30 per month | | |
|     4 > 30 per month (or chairbound) | rename old 4 into new 2 |     2 > 30 per month (or chairbound) |
| 6. Urinary incontinence | eliminate | |
|     0 None or a few drops less than daily | | |
|     1 A few drops staining clothes daily | | |
|     2 Large amounts, but only when asleep; no pad required during day | | |
|     3 Occasional large amounts in daytime; pad required | | |
|     4 Consistent, requiring diaper or catheter awake and asleep | | |
| 7. Sleep difficulty | eliminate | |
|     0 Neither 1º nor 2º insomnia (i.e., falls asleep easily and stays asleep) | | |
|     1 Either 1º or 2º insomnia; averages at least 5 hours sleep nightly | | |
|     2 Both 1º and 2º insomnia; averages at least 5 hours sleep nightly | | |
|     3 Either 1º or 2º insomnia; averages less than 5 hours sleep nightly | | |
|     4 Both 1º and 2º insomnia; averages less than 5 hours sleep nightly | | |
| **II. MENTAL EXAM** | | |
| 8. Disorientation | eliminate | |
|     0 Clearly absent | | |
|     1 Equivocal or minimal | | |
|     2 Clearly present, but not affecting activities of daily living (ADL) | | |
|     3 Interfering mildly with ADL | | |
|     4 Interfering markedly with ADL | | |
| 9. Bradyphrenia | eliminate | |
|     0 Clearly absent | | |
|     1 Equivocal or minimal | | |
|     2 Clearly present, but not affecting activities of daily living (ADL) | | |
|     3 Interfering mildly with ADL | | |
|     4 Interfering markedly with ADL | | |
| 10. Emotional incontinence | eliminate | |
|     0 Clearly absent | | |
|     1 Equivocal or minimal | | |
|     2 Clearly present, but not affecting activities of daily living (ADL) | | |
|     3 Interfering mildly with ADL | | |
|     4 Interfering markedly with ADL | | |
| 11. Grasping/imitatative/utilizing behavior | eliminate | |
|     0 Clearly absent | | |
|     1 Equivocal or minimal | | |
|     2 Clearly present, but not affecting activities of daily living (ADL) | | |
|     3 Interfering mildly with ADL | | |
|     4 Interfering markedly with ADL | | |
| **III. BULBAR EXAM** | | |
| 12. Dysarthria (ignoring palilalia) | | 12. Dysarthria (ignoring palilalia) |

| | | |
|---|---|---|
| 0 None | | 0 None |
| 1 Minimal; all or nearly all words easily comprehensible (to examiner, not family) | collapse old 1 and 2 into new 1 | 1 All or nearly all words comprehensible |
| 2 Definite, moderate; most words comprehensible | | |
| 3 Severe; may be fluent but most words incomprehensible | collapse old 3 and 4 into new 2 | 2 Severe; most words incomprehensible, verbal communication is ineffective |
| 4 Mute; or a few poorly comprehensible words | | |
| 13. Dysphagia (for 30-50 cc of water from a cup, if safe) | | 13. Dysphagia (for 30-50 cc of water from a cup, if safe) |
| 0 None | | 0 None |
| 1 Fluid pools in mouth or pharynx, or swallows slowly, but no choking/coughing | | 1 Single sips, or fluid pools in mouth or pharynx, but no choking/coughing |
| 2 Occasionally coughs to clear fluid; no frank aspiration | | 2 Occasionally coughs briefly to clear fluid or up to nectar thickened liquids |
| 3 Frequently coughs to clear fluid; may aspirate slightly; may expectorate frequently rather than swallow secretions | | 3 Frequently coughs forcefully to clear fluid may expectorate frequently ratherthan swallow secretions or honey or spoon thickened liquids |
| 4 Requires artificial measures (oral suctioning, tracheostomy or feeding gastrostomy) to avoid aspiration | | 4 Requires oral suctioning, tracheostomy, or feeding gastrostomy to avoidaspiration |

**IV. SUPRANUCLEAR OCULAR MOTOR EXAM**
Rate by inspection of saccades on command from the primary position of gaze to a stationary target.

| | |
|---|---|
| 14. Voluntary upward saccades | eliminate |
|     0 Not slow or hypometric; 86-100% of normal amplitude | |
|     1 Slow or hypometric; 86-100% of normal amplitude | |
|     2 51-85% of normal amplitude | |
|     3 16-50% of normal amplitude | |
|     4 15% of normal amplitude or worse | |
| 15. Voluntary downward saccades | eliminate |
|     0 Not slow or hypometric; 86-100% of normal amplitude | |
|     1 Slow or hypometric; 86-100% of normal amplitude | |
|     2 51-85% of normal amplitude | |
|     3 16-50% of normal amplitude | |
|     4 15% of normal amplitude or worse | |
| 16. Voluntary left and right saccades | eliminate |
|     0 Not slow or hypometric; 86-100% of normal amplitude | |
|     1 Slow or hypometric; 86-100% of normal amplitude | |
|     2 51-85% of normal amplitude | |
|     3 16-50% of normal amplitude | |
|     4 15% of normal amplitude or worse | |
| 17. Eyelid dysfunction | eliminate |
|     0 None | |
|     1 Blink rate decreased (< 15/minute) but no other abnormalities | |
|     2 Mild inhibition of opening or closing or mild blepharospasm; no visual disability | |
|     3 Moderate lid-opening inhibition or blepharospasm causing partial visual disability | |
|     4 Functional blindness or near-blindness because of involuntary eyelid closure | |

**V. LIMB EXAM**

| | |
|---|---|
| 18. Limb rigidity (rate the worst of the four) | eliminate |
|     0 Absent | |
|     1 Slight or detectable only on activation | |
|     2 Definitely abnormal, but full range of motion possible | |
|     3 Only partial range of motion possible | |
|     4 Little or no passive motion possible | |
| 19. Limb dystonia (rate worst of the four; ignore neck and face) | eliminate |
|     0 Absent | |
|     1 Subtle or present only when activated by other movement | |
|     2 Obvious but not continuous | |
|     3 Obvious but not continuous | |
|     4 Continuous and disabling | |
| 20. Finger tapping (if asymmetric, rate worse side) | eliminate |
|     0 Normal (> 14 taps/5 sec with maximal amplitude) | |
|     1 Impaired (6-14 taps/5 sec or moderate loss of amplitude | |
|     2 Barely able to perform (0-5 taps/5 sec or severe loss of amplitude) | |
| 21. Toe tapping (if asymmetric, rate worse side) | eliminate |
|     0 Normal (> 14 taps/5 sec with maximal amplitude) | |
|     1 Impaired (6-14 taps/5 sec or moderate loss of amplitude | |
|     2 Barely able to perform (0-5 taps/5 sec or severe loss of amplitude) | |
| 22. Apraxia of hand movement | eliminate |
|     0 Absent | |
|     1 Present, not impairing most functions | |
|     2 Impairing most functions | |
| 23. Tremor in any part | eliminate |
|     0 Absent | |
|     1 Present, not impairing most functions | |
|     2 Impairing most functions | |

**VI. GAIT/MIDLINE EXAM**

| | | |
|---|---|---|
| 24. Neck rigidity or dystonia | | 24. Neck rigidity or dystonia |
|     0 Absent | |     0 Absent |
|     1 Slight or detectable only when activated by other movement | collapse old 1 and 2 into new 1 |     1 Abnormal but full range of motion possible |
|     2 Definitely abnormal, but full range of motion possible | | |

| | | |
|---|---|---|
| 3 Only partial range of motion possible | rename old 3 into new 2 | 2 Only partial range of motion possible |
| 4 Little or no passive motion possible | rename old 4 into new 3 | 3 Little or no passive motion possible |
| **25. Arising from chair** | | **25. Arising from chair** |
| 0 Normal | collapse old 0, 1 and 2 into new 0 | 0 Normal may be slow or require more than 1 attempt, does not need to push off with hands |
| 1 Slow but arises on first attempt | | |
| 2 Requires more than one attempt, but arises without using hands | | |
| 3 Requires use of hands | rename old 3 into new 1 | 1 Requires use of hands |
| 4 Unable to arise without assistance | rename old 4 into new 2 | 2 Unable to arise without assistance |
| **26. Gait** | | **26. Gait** |
| 0 Normal | collapse old 0 and 1 into new 0 | 0 Normal base of support may be increased |
| 1 Slightly wide-based or irregular or slight pulsion on turns | | |
| 2 Must walk slowly or occasionally use walls or helper to avoid falling, especially on turns | rename old 2 into new 1 | 1 Must walk slowly or occasionally use walls or helper to avoid falling, especially on turns |
| 3 Must use assistance all or almost all the time | collapse old 3 and 4 into new 2 | 2 Must walk with assistance of a caregiver all or almost |
| 4 Unable to walk, even with walker; may be able to transfer | | |
| **27. Postural stability (on backward pull)** | | **27. Postural stability (on backward pull)** |
| 0 Normal (shifts neither foot or one foot) | collapse old 0 and 1 into new 0 | 0 Normal, may reposition both feet but maintains balance unaided |
| 1 Must shift each foot at least once but recovers unaided | | |
| 2 Shifts feet and must be caught by examiner | rename old 2 into new 1 | 1 Shifts feet and must be caught by examiner |
| 3 Unable to shift feet; must be caught, but does not require assistance to stand still | rename old 3 into new 2 | 2 Unable to shift feet; must be caught, but does not require assistance to stand still |
| 4 Tends to fall without a pull; requires assistance to stand still | rename old 4 into new 3 | 3 Tends to fall without a pull; requires assistance to stand still |
| **28. Sitting down (may touch seat or back but not arms of chair)** | | **28. Sitting down (may touch seat or back but not arms of chair)** |
| 0 Normal | collapse old 0 and 1 into new 0 | 0 normal or slightly awkward, positions self in front of chair, controlled descent into chair |
| 1 Slightly stiff or awkward | | |
| 2 Easily positions self before chair, but descent into chair is uncontrolled | rename old 2 into new 1 | 1 Easily positions self before chair, but descent into chair is uncontrolled |
| 3 Has difficulty finding chair behind him/her and descent is uncontrolled | rename old 3 into new 2 | 2 Has difficulty finding chair behind him/her and descent is uncontrolled |
| 4 Unable to test because of severe postural instability | rename old 4 into new 3 | 3 Unable to sit without assistance |



\end{document}